%% file: main.tex
\newcommand{\PaperTitle}{FP-Agent: Fingerprinting AI Browsing Agents}

\documentclass[10pt,sigconf,letterpaper,nonacm]{acmart}

\author{Ethan Wang}
\orcid{0009-0000-2715-0410}
\affiliation{
  \institution{\textit{University of California, Davis}}
  \country{}
}
\email{ebwang@ucdavis.edu}

\author{Zubair Shafiq}
\affiliation{
  \institution{\textit{University of California, Davis}}
  \country{}
}
\email{zubair@ucdavis.edu}

\author{Yash Vekaria}
\affiliation{
  \institution{\textit{University of California, Davis}}
  \country{}
}
\email{yvekaria@ucdavis.edu}

\usepackage{graphicx}
\usepackage{soul}
\usepackage{multirow}
\usepackage{xcolor}
\usepackage[table]{xcolor}
\usepackage{tcolorbox}

\usepackage{makecell}

\newcommand{\aibot}{browsing agent}
\newcommand{\uppercaseaibot}{Browsing Agent}
\newcommand{\capitalizeaibot}{Browsing agent}
\newcommand{\para}[2][\relax]{{#1}\noindent \textbf{#2}.}

\definecolor{lightgreen}{RGB}{180, 220, 180}  
\definecolor{lightred}{RGB}{220, 180, 180}    
\definecolor{mousecolor}{HTML}{DCB4B4}
\definecolor{typecolor}{HTML}{DCC4A0}
\definecolor{scrollcolor}{HTML}{C4B4DC}

\newcommand{\tagpad}{\rule[-3pt]{0pt}{11pt}}  
\newcommand{\tagmouse}[1]{\colorbox{mousecolor}{\tagpad#1}}
\newcommand{\tagtype}[1]{\colorbox{typecolor}{\tagpad#1}}
\newcommand{\tagscroll}[1]{\colorbox{scrollcolor}{\tagpad#1}}

\tcbset{
  takeawaybox/.style={
    colback=gray!10,
    colframe=black,
    fonttitle=\bfseries,
    left=2pt, right=2pt, top=1pt, bottom=1pt,
    boxrule=0.5pt
  }
}

\tcbset{
    promptbox/.style={
      colback=gray!5,
      colbacktitle=gray!20,
      coltitle=black,
      fonttitle=\bfseries\small,
      arc=4pt,
    }
}

\newcommand{\macosicon}{\includegraphics[height=\dimexpr0.9\ht\strutbox+0.9\dp\strutbox\relax]{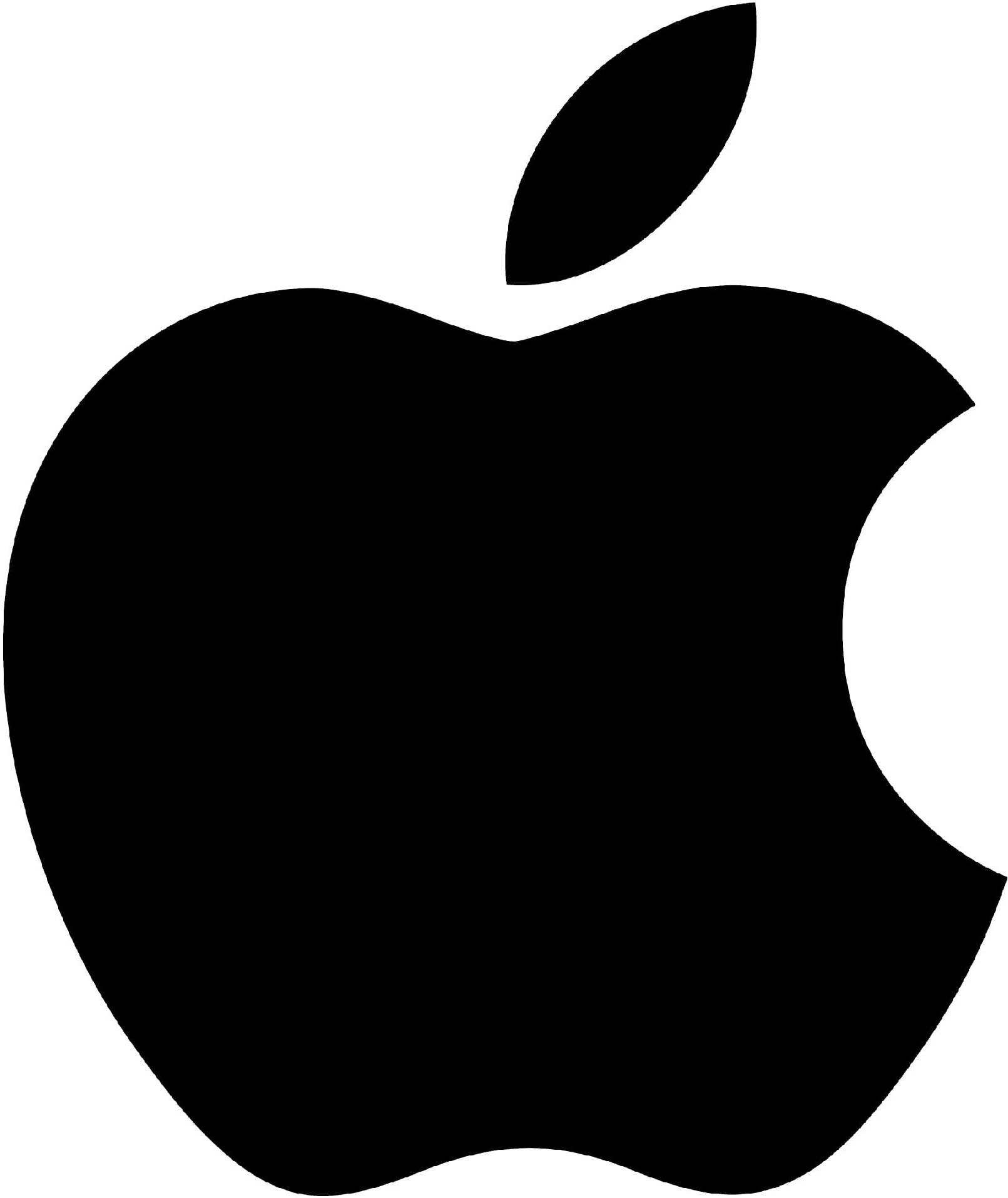}}
\newcommand{\windowsicon}{\includegraphics[height=\dimexpr0.9\ht\strutbox+0.9\dp\strutbox\relax]{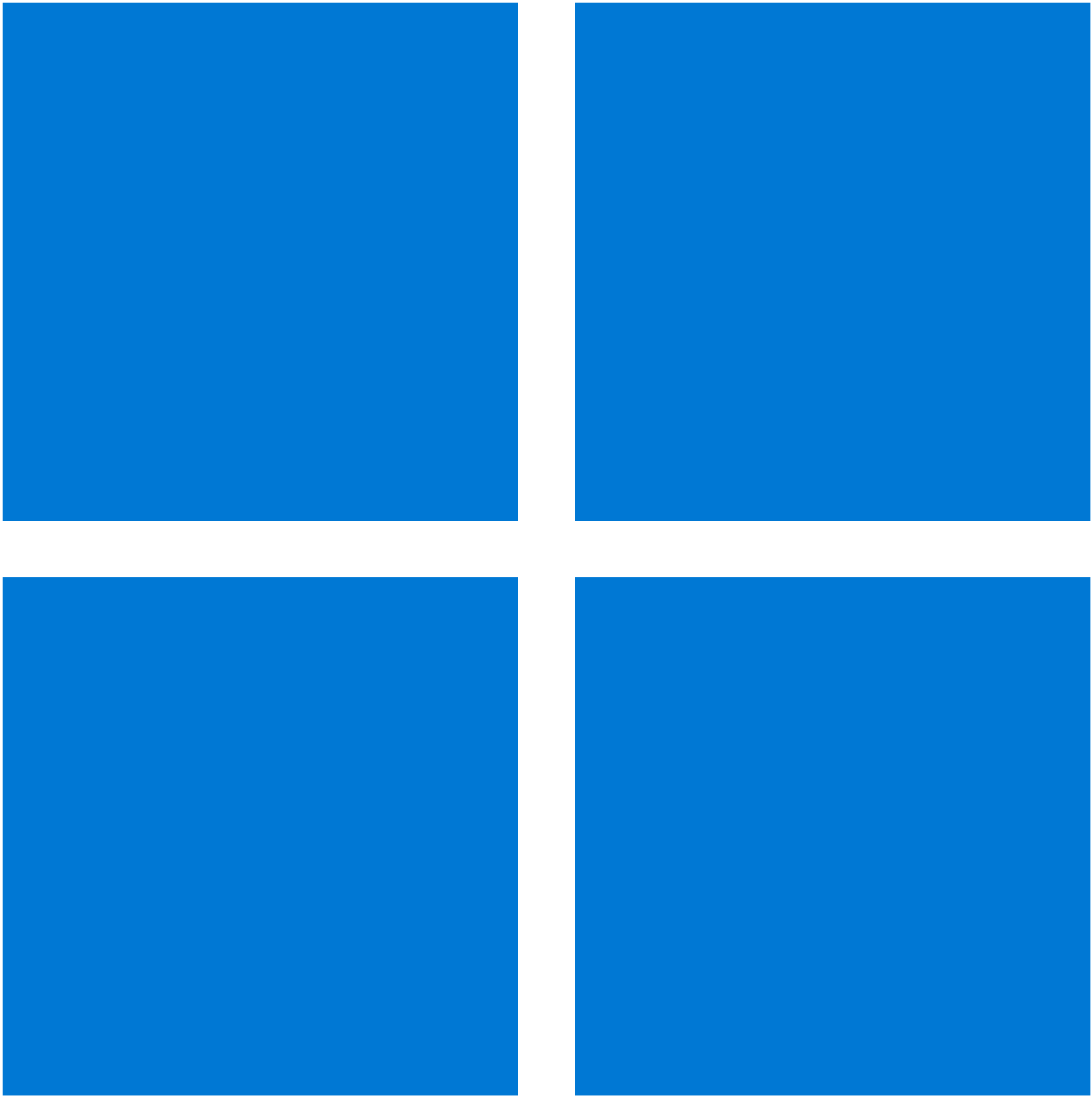}}
\newcommand{\linuxicon}{\includegraphics[height=\dimexpr0.9\ht\strutbox+0.9\dp\strutbox\relax]{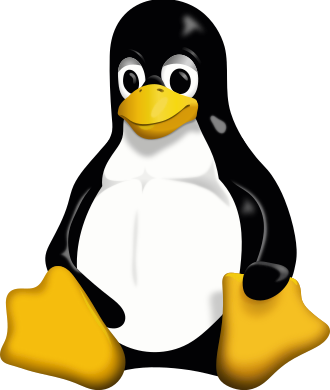}}
\newcommand{\atlasicon}{\includegraphics[height=\dimexpr0.9\ht\strutbox+0.9\dp\strutbox\relax]{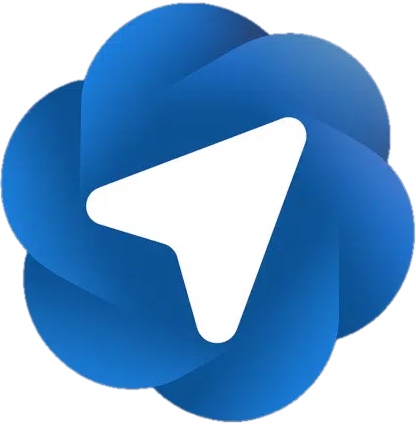}}
\newcommand{\browseruseicon}{\includegraphics[height=\dimexpr0.9\ht\strutbox+0.9\dp\strutbox\relax]{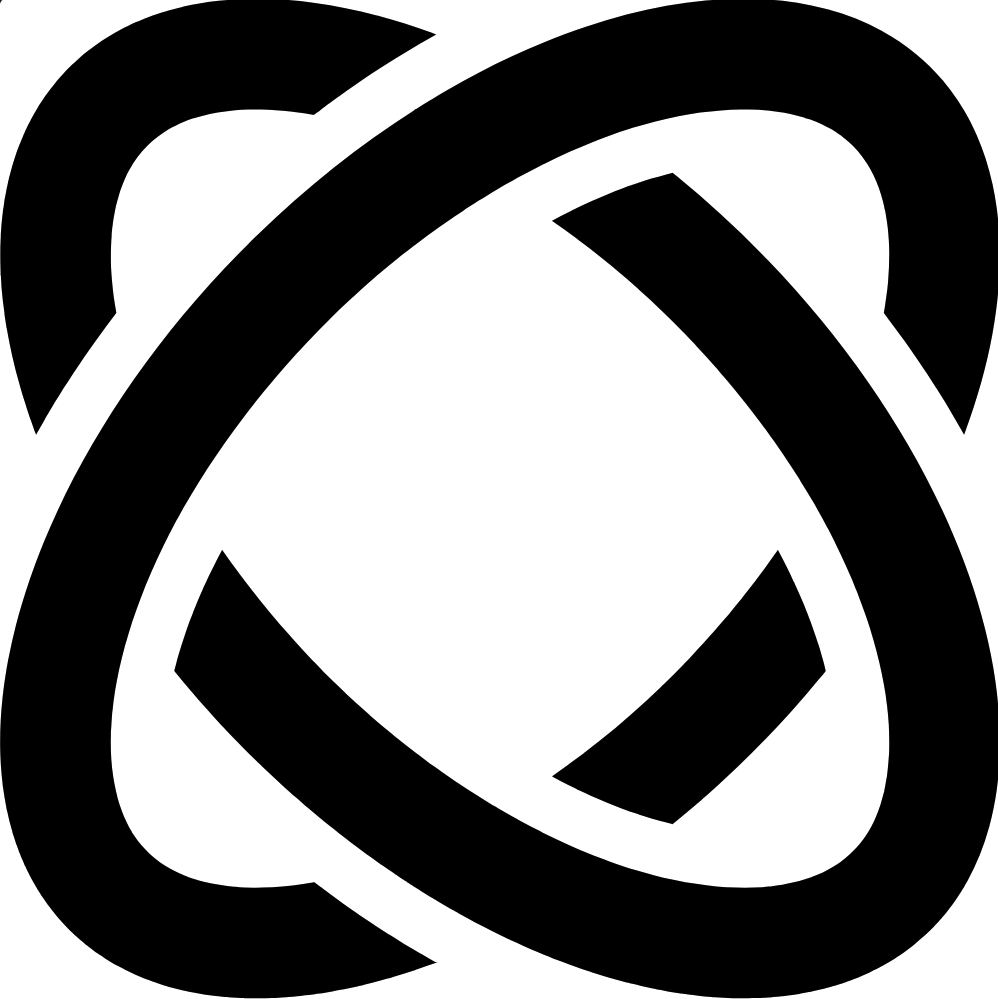}}
\newcommand{\chatgpticon}{\includegraphics[height=\dimexpr0.9\ht\strutbox+0.9\dp\strutbox\relax]{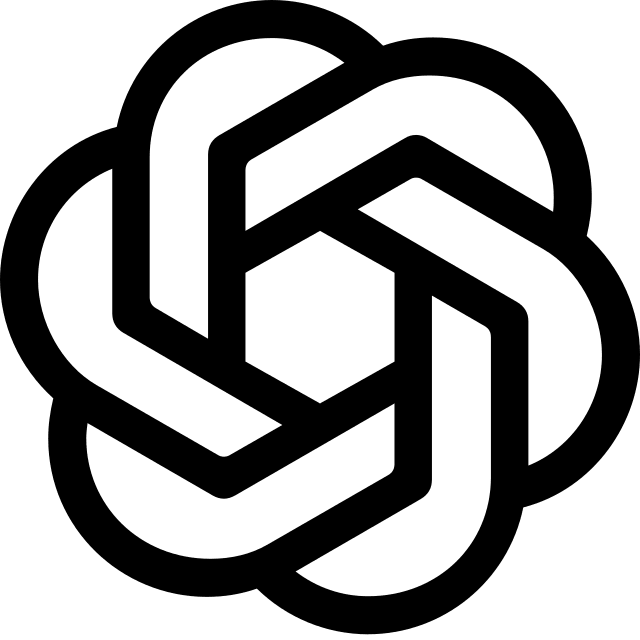}}
\newcommand{\claudeicon}{\includegraphics[height=\dimexpr0.9\ht\strutbox+0.9\dp\strutbox\relax]{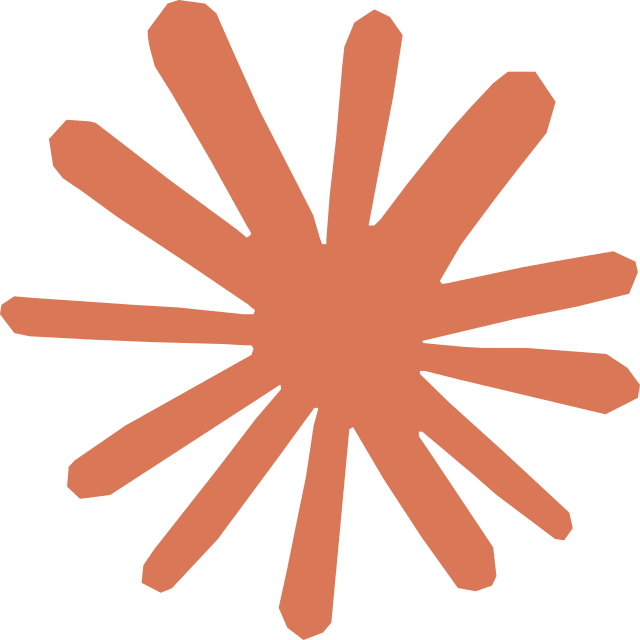}}
\newcommand{\cometicon}{\includegraphics[height=\dimexpr0.9\ht\strutbox+0.9\dp\strutbox\relax]{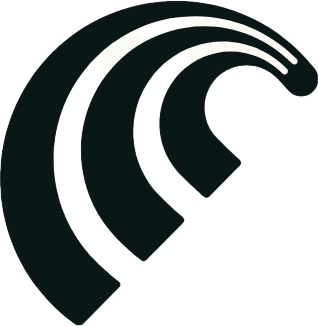}}
\newcommand{\manusicon}{\includegraphics[height=\dimexpr0.9\ht\strutbox+0.9\dp\strutbox\relax]{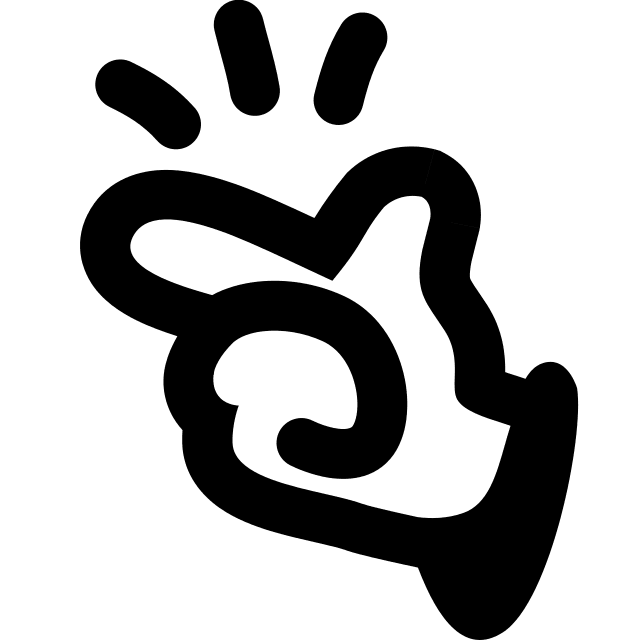}}
\newcommand{\skyvernicon}{\includegraphics[height=\dimexpr0.9\ht\strutbox+0.9\dp\strutbox\relax]{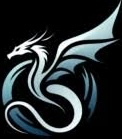}}
\newcommand{\humanicon}{\includegraphics[height=\dimexpr0.9\ht\strutbox+0.9\dp\strutbox\relax]{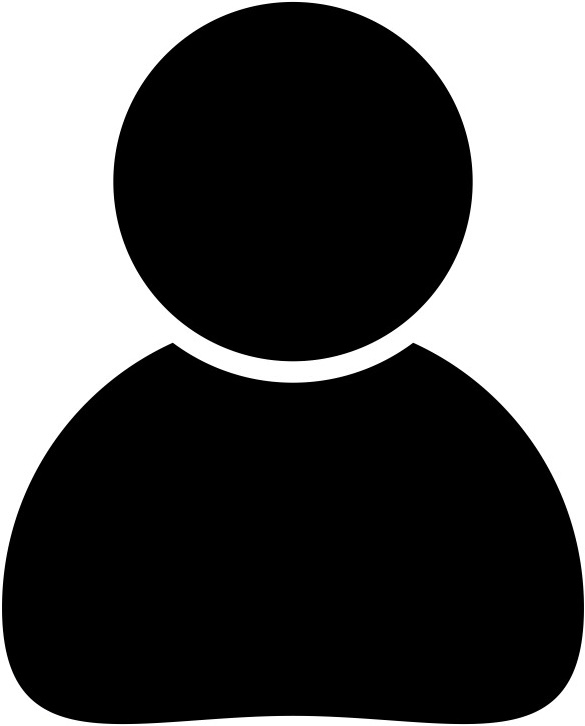}}
\newcommand{\githubstaricon}{\includegraphics[height=\dimexpr0.9\ht\strutbox+0.9\dp\strutbox\relax]{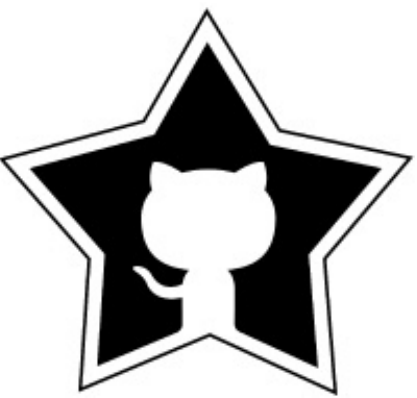}}

\begin{document}

\title{\PaperTitle}

\begin{abstract}
\input{sections/00_abstract}
\end{abstract}

\maketitle

\input{sections/01_introduction}

\input{sections/02_background}
\input{sections/03_related_work}
\input{sections/04_methodology}
\input{sections/05_characterizing_fingerprints}

\input{sections/06_discussion_and_limitations}
\input{sections/07_conclusion}

\begin{acks}

This work was supported in part by the National Science Foundation under award numbers 2138139 and 2103439.

\end{acks}

\input{sections/08_ethics}
\input{sections/09_data_availability}


\newpage
\bibliographystyle{ACM-Reference-Format}
\bibliography{main}

\appendix

\newpage
\onecolumn
\input{sections/10_task_details}
\newpage
\input{sections/11_task_screenshots}
\clearpage
\input{sections/12_behavioral_features}
\newpage
\input{sections/13_confusion_matrix}
\input{sections/14_keystroke_dynamics}

\end{document}

%% file: sections/00_abstract.tex
AI browsing agents are an emerging class of AI-powered bots capable of autonomously navigating websites. 
Unlike traditional web bots, AI browsing agents typically operate using real browsers and perform everyday tasks, making them difficult to detect.
Yet little is known about whether existing AI browsing agents can be distinguished from humans and one another based on their browser or behavioral fingerprints.
In this paper, we present the first controlled measurement study of seven AI browsing agents and human users.
Using an instrumented honey website, we collect browser and behavioral fingerprint features while AI browsing agents and humans perform three tasks: flight booking, online shopping, and forum interaction.
We then train \texttt{FP-Agent}, a multi-class classifier, to evaluate the discriminative power of these features.
We find that browser fingerprints provide limited discriminative power when shared by multiple AI browsing agents.
Behavioral fingerprints, however, are distinctive: differences in typing, scrolling, and mouse behavior separate AI browsing agents from humans and one another.
In a case study evaluating Cloudflare’s bot detection, \texttt{FP-Agent} detects all seven AI browsing agents, whereas Cloudflare detects only one.
Our findings show that behavioral fingerprints are a critical component to reliably detect and control this emerging form of web traffic.

%% file: sections/01_introduction.tex
\section{Introduction}
\label{sec:intro}

AI {\aibot}s (hereafter referred to as {\aibot}s) are an emerging type of web bot that leverage AI to autonomously browse websites on behalf of users.
Unlike traditional web bots, browsing agents use real browsers and perform tasks typically done by humans.
These properties make them substantially harder to distinguish from human traffic using conventional signals.
At the same time, modern bot detection remains an arms race~\cite{lee2025building, cresci2017paradigm} that increasingly relies on observable browser~\cite{human2026botdetection} and behavioral~\cite{kaiserman2026aiagent} fingerprint features to separate bots from humans, while adversaries continually adapt by manipulating these features to mimic humans~\cite{lee2025building, cresci2017paradigm, human2026botdetection}.

This challenge matters now because AI-driven web automation is becoming increasingly prevalent~\cite{human2026aitrafficreport}.
Cloudflare reported that 38.7\% of the top-million websites received AI bot traffic in 2024~\cite{bocharov2024aindependence}; by 2025, AI bots accounted for 8.7\% of all HTML traffic~\cite{cloudflare2025yearreview}.
Much of this traffic is driven by benign use cases, such as AI chatbots performing real-time crawls to search the web~\cite{openai2024search, anthropic2025search} and browsing agents performing web tasks on behalf of users~\cite{openai2026chatgptagent, levine2025agenticcommerce}.
Yet the same capabilities can also exacerbate malicious automation, including CAPTCHA solving, large-scale scraping, credential stuffing, and scalping~\cite{imperva2025badbot, akamai2025fraud, akamai2026navigating, cloudflare2025yearreview, levine2025agenticcommerce}.
For website publishers, distinguishing browsing-agent traffic from human traffic is therefore important not only for security, but also for access control, traffic analytics, infrastructure optimization, and emerging billing models such as pay per crawl~\cite{allen2025paypercrawl, akamai2026navigating, martinho2026markdown}.

Existing defenses, however, are often insufficient for this setting.
Current publisher-facing mechanisms largely rely on self-declaration: websites can express preferences through robots.txt directives~\cite{koster2022robotsexclusion}, maintain allowlists of known AI traffic~\cite{cloudflare2026botmanagebrief, human2026botdetection}, or use emerging mechanisms such as Web Bot Auth to authenticate bot identity~\cite{meunier2026webbotauth}.
These approaches are useful when adopted, but they remain fundamentally voluntary.
Malicious operators can simply avoid identifying themselves, and even benign ones may not adopt or use these mechanisms consistently~\cite{segura2026identity, cloudflare2026botdirectory}.
As a result, self-identification alone is not a reliable basis for detecting and controlling browsing-agent traffic in practice.

What remains unclear is whether the current generation of browsing agents nevertheless exhibit properties that distinguish them from humans and from one another.
Despite the growing prevalence of browsing agents, little is known about their browser- and interaction-level behavior during realistic web tasks.
In particular, it is unknown whether browser fingerprints and behavioral fingerprints provide sufficient discriminative power for practical multi-class classification of browsing agents and humans, and how these compare to existing deployed defenses.
To the best of our knowledge, no prior work has systematically measured these fingerprint features for {\aibot}s in a controlled setting.

In this paper, we address this gap by conducting a measurement study of seven {\aibot}s and human users.
We deploy an instrumented honey website to collect browser and behavioral fingerprint features while {\aibot}s and humans perform three tasks: flight booking, online shopping, and forum interaction.
These tasks involve exercising common web interaction primitives under controlled conditions, allowing us to characterize how {\aibot}s and humans differ in the collected features.
We then train \texttt{FP-Agent}, a multi-class classifier, to evaluate the discriminative power of these fingerprint features.
Finally, we compare \texttt{FP-Agent} against Cloudflare's bot detection.

Overall, our measurement study yields three key findings. 
First, browser fingerprints alone provide limited discriminative power, especially when multiple {\aibot}s share the same browser fingerprint.
Second, behavioral fingerprints are highly distinctive: typing, scrolling, and mouse behavior reliably distinguishes {\aibot}s from humans and from one another.
These differences include concrete browsing-agent-specific quirks, such as paste-based versus keystroke-based typing, change-event-based form filling, repeated delete-and-retype patterns, direct jumps to click targets without continuous mouse movement, and distinct scrolling styles ranging from recurring instantaneous jumps to discrete multi-burst exploration.
Third, these behavioral fingerprint features enable accurate multi-class classification and expose limitations of current deployed defenses: in our Cloudflare case study, \texttt{FP-Agent} detects all seven studied agents, whereas Cloudflare blocks only one.

Our work highlights the limits of current defenses and shows the promise of behavioral fingerprinting for detecting and controlling {\aibot}s today. 
We find that behavioral fingerprint features enable accurate multi-class classification and short-window real-time detection.
However, our findings are limited by coverage of {\aibot}s, tasks, and system configurations, and future {\aibot}s may adopt more human-like behaviors. 
\texttt{FP-Agent} provides an extensible measurement framework for future research on measuring and characterizing traffic of {\aibot}s.

%% file: sections/02_background.tex
\section{Background}

\subsection{AI Bots}
\label{sec:ai_bots}

\para{Traditional bots}
Web bots, programs used to automate web interactions~\cite{gray1993credits, mirtaheri2013brief}, comprised 51\% of all web traffic in 2024~\cite{imperva2025badbot}.
Although useful for a variety of essential tasks such as indexing webpages and performing web measurements~\cite{cloudflare2026whataregoodbots}, web bots can also be used for malicious activities such as spam, fraud, denial of service (DoS), and invalid traffic (IVT)~\cite{cloudflare2026whataregoodbots, imperva2025badbot, human2026invalidtraffic}.
Up to 37\% of bot traffic is reported to be malicious~\cite{imperva2025badbot}, necessitating monitoring and defending against their continuously evolving behavior.

\para{Emergence of AI bots}
According to Cloudflare, 38.7\% of web traffic in 2024 originated from AI bots~\cite{bocharov2024aindependence}.
These AI web bots can be categorized as AI training crawlers, AI search crawlers, AI assistant crawlers, and AI {\aibot}s.
\textit{\textbf{AI training crawlers}} (e.g., OpenAI's GPTBot~\cite{openai2026overview}) crawl the web to collect data for training LLMs.
\textit{\textbf{AI search crawlers}} (e.g., Anthropic's Claude-SearchBot~\cite{anthropic2026doesanthropiccrawl}) crawl the web and create a static index for AI-powered search.
Companies claim that the data scraped by AI search crawlers is not used for training their models~\cite{liu2025somesite}.
\textit{\textbf{AI assistant crawlers}} (e.g., Perplexity's Perplexity-User~\cite{perplexity2026crawlers}) perform on-demand crawls to retrieve data for response augmentation.
A fourth emerging 

\noindent category called \textit{\textbf{AI {\aibot}s}}\footnote{{\capitalizeaibot}s have also been termed ``browser agents''~\cite{ukani2025privacy} and ``web agents''~\cite{li2026webcloak} in prior literature.} (e.g., Manus Bot~\cite{cloudflare2026manusbot}) 

\noindent leverage LLMs to autonomously perform web-browsing tasks.
AI {\aibot}s can be run in the cloud or locally using browser extensions (e.g., Claude for Chrome~\cite{chrome2026claudeextension}) or agentic browsers (e.g., ChatGPT Atlas~\cite{openai2026atlas}).
In this paper, we study \textit{AI {\aibot}s}.

\para{{\capitalizeaibot} architectures}
A typical {\aibot} implementation involves an underlying LLM model for reasoning and planning, a set of tools that the browsing agent can call to interact with a website, and a system prompt to orchestrate how the browsing agent acts~\cite{deng2023mind2web, he2024webvoyager, openai2026practicalguide}.
This is operationalized as either a single agent or as part of a multi-agent architecture in which specialized sub-agents handle distinct sub-tasks \cite{masterman2024landscape}.
Depending on how the browser state is conveyed to the LLM, {\aibot}s can be categorized as DOM-based, vision-based, or hybrid \cite{deng2023mind2web, unclecode2024crawl4ai, he2024webvoyager, yang2023set}: DOM-based browsing agents parse a website's Document Object Model to understand its structure, vision-based browsing agents rely on annotated screenshots, and hybrid ones combine both representations.
At each step of the browsing-agent control loop, the LLM outputs tool calls based on the browser state at that point, which are then executed by the browsing agent.
These tools are packaged as wrappers around either a browser automation framework (e.g., Playwright\cite{playwright2026playwrightrepo}) or the Chrome DevTools Protocol (CDP)~\cite{google2026chromedevtools} itself.

\para{Impact of AI bots}
Website publishers have raised concerns about copyright infringement and monetary loss stemming from AI bots consuming their content without driving human traffic \cite{liu2025somesite, hadero2023nytsuesopenai, obrien2025redditsuesperplexity, weil2025hiddeneconomics}.
Solutions like Cloudflare's pay per crawl~\cite{allen2025paypercrawl} attempt to mitigate monetary concerns, but still require AI bots to be detected for billing.
The influx of AI bot traffic may also cause website accessibility issues for humans by consuming server bandwidth with frequent crawling~\cite{duke2026impact}.
Furthermore, with reasoning capabilities, {\aibot}s can perform more sophisticated attacks~\cite{imperva2025badbot}, posing new security risks.
{\capitalizeaibot}s may also fall victims to prompt injection attacks orchestrated by malicious websites to extract their user's sensitive information~\cite{openai2025promptinjection, perplexity2025promptinjection, brave2025agenticbrowsersecurity}.

\subsection{Defending Against AI Bots}
\label{sec:defenses_against_ai_bots}

\para{Industry solutions}
Websites can include a robots.txt~\cite{koster2022robotsexclusion} file at their root to indicate crawling rules for different user agents.
Recently, Cloudflare also introduced Content Signals~\cite{cloudflare2026contentsignals}, creating additional fine-grained directives to specify whether AI bots can reference, index, or train using the website's content.
Moreover, web application firewalls (WAFs) can rate-limit or block requests classified as AI bot traffic~\cite{cloudflare2026botmanagebrief, human2026botdetection}.
Industrial solutions---such as Cloudflare's Bot Management~\cite{cloudflare2026botmanagebrief}, HUMAN's Agentic Trust~\cite{human2026agentictrust}, and Data- 

\noindent Dome's Agent Trust~\cite{datadome2026agenttrust}---have also been developed, employing a combination of rules, patterns, challenges, honeypots, and machine learning techniques to detect AI bot traffic. 

Self-identification is also common in AI bot detection.
AI bots may publicize their user agent and IP range~\cite{cloudflare2026botdirectory, perplexity2026crawlers, openai2026overview} or enroll in bot registration programs (e.g., Cloudflare's verified bots~\cite{cloudflare2026verifiedbots}).
Bot detection companies have also begun implementing Web Bot Auth~\cite{meunier2026webbotauth}, an emerging mechanism built on HTTP message signatures~\cite{backman2024httpmessage} that allows AI bots to declare their identity by cryptographically signing part of their requests.
Such cryptographic signatures are harder to spoof and provide stronger guarantees of identity.

\para{Challenges in {\aibot} detection}
AI bots may conditionally respect robots.txt~\cite{kim2025scrapers} or ignore it entirely~\cite{liu2025somesite, cui2025odyssey, kim2025scrapers}.
It can also be frequently misconfigured~\cite{cui2025odyssey} and misinterpreted by AI bots~\cite{liu2025somesite} due to its complexity.
Additionally, while some AI bots can be straightforward to detect, techniques such as modifying and shuffling IP addresses and browser fingerprints~\cite{imperva2025badbot}, keeping traffic rates low~\cite{cloudflare2026lowslow}, and mimicking human-like behavioral patterns have proved to be effective in evading basic bot-detection systems~\cite{human2026botdetection}.

{\capitalizeaibot}s pose additional challenges.
First, they can utilize their underlying LLM's reasoning capabilities to solve challenges (e.g., CAPTCHA) and handle dynamic websites \cite{imperva2025badbot}.
Second, since many {\aibot}s use a Chromium-based browser, their user agent strings are generic \cite{human2026comet, human2026atlas, human2026chatgptagent}.
Moreover, some are operated locally, producing network traffic that originates from residential networks \cite{human2026comet, human2026atlas}.
These characteristics make them indistinguishable from humans based on IP address and user agent string~\cite{human2026comet, human2026atlas, human2026chatgptagent}.
Third, not all {\aibot}s identify themselves \cite{human2026atlas, human2026comet}, rendering voluntary systems such as Web Bot Auth ineffective.

%% file: sections/03_related_work.tex
\section{Related Work}

\subsection{Bot Measurement Studies}
\label{sec:bot_measurement}

Bot characterization has been studied extensively across a range of deployments on the web.
Prior work has leveraged honeypot networks to examine the request behavior alongside network-level signals of web bots \cite{kondracki2022uninvited, li2021good} and to identify SSH bot attack behaviors and strategies~\cite{munteanu2025attacks}.
Recently, Venugopalan et al. \cite{venugopalan2025fp} used a similar approach to characterize inconsistent browser fingerprint attributes used by evasive bots in the wild.
Other studies~\cite{na2023evolving, gianvecchio2008measurement} have extended bot measurements to social media, examining their activity frequency, comments, and interaction behaviors.
More recent studies have focused on understanding security and privacy of AI bots~\cite{ukani2025privacy, roesner2026agentic, liu2025somesite} and evaluating blanket defenses such as cloaking against AI bots~\cite{li2026webcloak, ayzenshteyn2025cloak}.

\subsection{Web Bot Detection}
\label{sec:web_bot_detection}
\textbf{Browser fingerprints} are widely used as a differentiating 

\noindent signal between bots and humans~\cite{venugopalan2025fp, vastel2020fp, amin2020web, llamas2025balancing}.
Browser and system characteristics (e.g., plugins, screen resolution, timezone) exposed through web APIs and the rendering of on-page elements can be combined into a stable fingerprint~\cite{amin2020web}.
Vastel et al. found that 31.96\% of the Alexa top 10K websites that block crawlers use browser fingerprints \cite{vastel2020fp}.
Prior works have commonly employed machine learning to learn patterns derived from browser fingerprints~\cite{venugopalan2025fp, amin2020web}.

\para{Behavioral fingerprints} Studies have evaluated the efficacy of features derived from browsing behaviors (e.g., request sequences~\cite{cabri2018online}, request paths and timestamps~\cite{herley2022automated}) for detection using statistical methods and machine learning classifiers.
Behavioral biometrics from website interactions, such as keystroke latencies and mouse movement patterns, have also been shown to reliably distinguish bots from humans~\cite{chu2013blog, ousat2024breaking, niu2021mouse}.
In addition to behavioral fingerprints obtained from real bot traffic, recent works have evaluated the use of synthetic mouse movements~\cite{acien2022becaptcha} and keystrokes~\cite{dealcala2023becaptcha, see2024detecting} to train classifiers to detect real bots.

\subsection{Distinction from Prior Work}
\label{sec:disctinction_from_prior_work}

To the best of our knowledge, no work has systematically studied the browser fingerprint characteristics and website interaction behaviors of {\aibot}s nor their differentiability.
Our work addresses this gap by characterizing browser and behavioral fingerprints of seven proprietary and open-source {\aibot}s and evaluating its distinguishing power for {\aibot} detection regardless of intent, providing empirical measurements to enable fine-grained controls for website operators.

%% file: sections/04_methodology.tex
\section{Methodology}
\label{sec:methodology}

\begin{figure}
      \centering
      \includegraphics[width=0.99\linewidth]{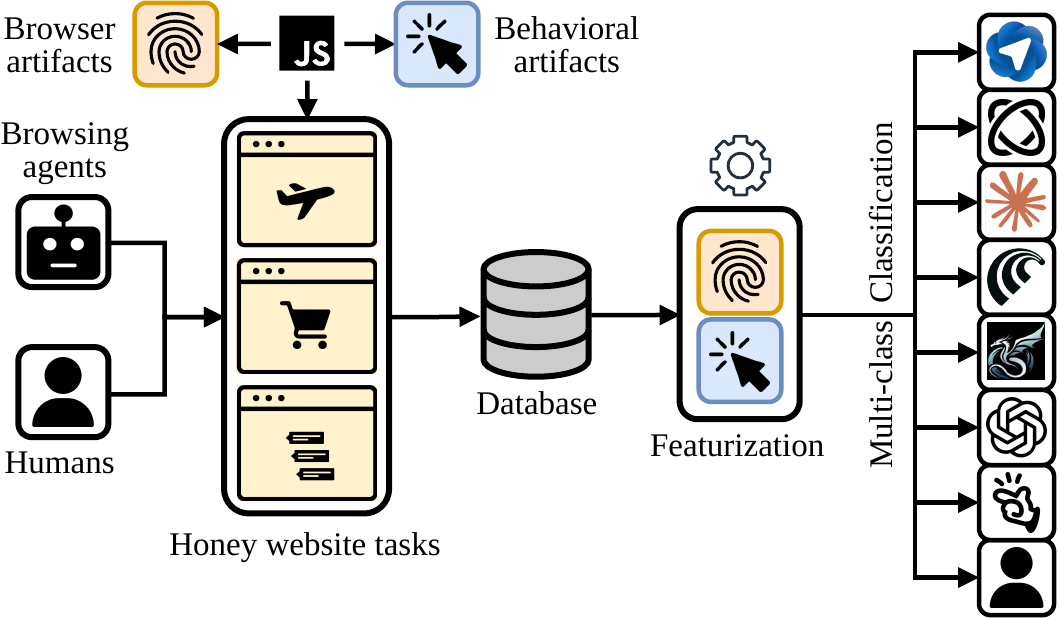}
      \vspace{-2.5mm}
      \caption{An overview of our \texttt{FP-Agent} framework.}
      \Description[]{An overview of our \texttt{FP-Agent} framework.}
      \label{fig:methodology}
      \vspace{-4mm}
\end{figure}

In this section, we describe our measurement methodology to systematically characterize and fingerprint {\aibot}s.
Figure \ref{fig:methodology} provides an overview of our methodology, which includes {\aibot} selection, honey website instrumentation, task execution, data collection, featurization, and multi-class classification.
We make \textit{all} artifacts of our work \textit{fully} available to the public at \textit{\url{https://github.com/ethanbwang/fp-agent}} (see Appendix~\ref{appendix:data_availability}).

\subsection{Selection of {\uppercaseaibot}s}
\label{sec:browsing_agent_selection}

We select five {\aibot}s offered by leading AI companies~\cite{stanford2025aiindexreport}: OpenAI's \textit{Atlas Browser} \cite{openai2026atlas} and \textit{ChatGPT Agent} \cite{openai2026chatgptagent}, Anthropic's \textit{Claude for Chrome} \cite{chrome2026claudeextension}, Perplexity's \textit{Comet} \cite{perplexity2026comet}, and Meta's \textit{Manus} \cite{meta2026manus}.
These include the most prominent proprietary {\aibot}s available to end users today and capture diversity in deployment model and platform support. 
Google's {\aibot} \textit{Project Mariner} was in its beta stage during our selection period and could not complete the tasks, resulting in its exclusion, while xAI did not have a {\aibot}.
To complement the proprietary systems, we also include two well-known open-source {\aibot}s, \textit{Browser Use} \cite{browseruse2026browseruserepo} and \textit{Skyvern} \cite{skyvern2026skyvernrepo}, allowing us to compare commercial and open-source browsing agent implementations. 
For reproducibility, we fix the versions of Skyvern and Browser Use to 0.2.23 and 0.9.2 respectively.
Table \ref{tab:fpjs_table} summarizes the seven selected {\aibot}s, including their user base size, deployment model, and system coverage.

\subsection{Honey Website Infrastructure}
\label{sec:honey_website_design}

\para{Website design}
We design and deploy a controlled, instrumented honey website to study browsing patterns.
To isolate each visitor's browsing patterns, we map observed traffic to the specific visitor that generated it.
To this end, following prior work \cite{venugopalan2025fp}, we create visitor-specific versions of the website as subpages identified by a 10-character random path string.
Only requests to valid subpages are served by the honey website; all other requests receive a 404 response, preventing noise from unsolicited visits by random web crawlers or humans.
This design enables the automatic deployment of distinct website versions for each {\aibot} and human participant, providing reliable ground truth for our study.

\para{Client-side instrumentation}
We implement a client-side JavaScript instrumentation script to collect the browser and behavioral fingerprint features produced by each visitor, which we collectively refer to as ``artifacts.'' 
The script leverages browser-exposed Web APIs to capture browser and system characteristics and record user actions on the page.
To collect browser fingerprints, we use FingerprintJS \cite{fingerprint2026fingerprintjs}, a widely used open-source browser fingerprinting library.
To collect behavioral fingerprints, we implement custom JavaScript event listeners that capture website interaction events.
The collected artifacts are sent from the client-side honey website to our web server and stored in a database.

\subsection{\textbf{Tasks}}
\label{sec:tasks}

\para{Task Descriptions}
To elicit representative browsing behavior, we design three tasks that capture common but diverse web interaction patterns: booking a flight, online shopping, and interacting with community forums.
Rather than modeling specific websites, we construct task-specific subpages that systematically exercise core interaction primitives (text input, scrolling, pointing, and clicking) that are broadly used across modern web applications.
This design allows us to observe how {\aibot}s and humans perform similar interaction types under controlled conditions.
The interactions targeted by each task are listed in Table~\ref{tab:task_interactions}.

The \textit{\textbf{flight-booking task}} models a multi-step form workflow typical of travel websites, and includes short text inputs, date pickers, radio buttons, drop-down menus, toggle controls, and scrolling within elements (see Figure~\ref{fig:flight_booking_and_shopping_screenshots}).
The \textit{\textbf{shopping task}} models a standard e-commerce workflow, involving search, filtering, item selection, and cart interaction (see Figure~\ref{fig:flight_booking_and_shopping_screenshots}).
The \textit{\textbf{forums task}} models a community discussion workflow, involving navigation to a thread, reading via scrolling, and composing a long-form text response (see Figure~\ref{fig:forums_screenshots}).
The prompt for each task is included in Appendix~\ref{appendix:prompts} and screenshots of task pages are provided in Appendix~\ref{appendix:task_screenshots}.

\para{{\capitalizeaibot} performance on tasks}
All {\aibot}s completed all tasks, typically within three minutes.
The flight-booking task took the longest, whereas the shopping and forums tasks required similar completion times.
Skyvern was consistently slower than the other {\aibot}s, taking up to 10 minutes on the flight-booking task.

\vspace{-1.3mm}
\subsection{Data Collection}
\label{sec:data_collection}

\para{{\capitalizeaibot} data collection}
To collect artifacts, we run 1000 trials in total for each {\aibot} equally split across the three tasks, with each trial executed on a unique task- and agent-specific subpage of our honey website.
For {\aibot}s that support multiple systems (Table~\ref{tab:fpjs_table}), we distribute trials evenly across supported systems to capture system-specific variation in the collected artifacts.
For local tests, we use an M1 MacBook Air running macOS 26.2, an Ubuntu 24.04 desktop, and a Windows 10 laptop.
Despite supporting Linux, we were unable to automate data collection for Claude on the Ubuntu system due to the GUI automation library we use not supporting Wayland~\cite{kavyanshkhaitan2025waylandsupport}.

\para{Human data collection}
To collect human artifacts, we recruit 56 undergraduate students from our university to participate in our study.
Each participant performs three repetitions of each task.
To capture representative browser fingerprints, participants complete the tasks on their own systems.
To protect participant privacy, we ensure that collected data cannot be linked back to individuals, for example by hashing IP addresses before storage.
Our study protocol was approved by our university's IRB (see Appendix~\ref{sec:ethics}).

\subsection{Featurization}
\label{sec:features}
We process the collected artifacts into 418 browser finger- 

\noindent print and 50 behavior fingerprint features.
The browser fingerprint features are derived from a subset of the attributes collected by FingerprintJS.
We one-hot encode categorical attributes and represent numerical attributes as floating-point values.
The behavioral features are organized by browser interaction type, as described in Section~\ref{sec:tasks}.
To characterize mouse behavior, we use the screen-resolution-agnostic mouse movement features from \cite{zheng2016efficient}; to characterize typing behavior, we use inter-key and hold latencies as defined in~\cite{acien2021typenet}.
We also include boolean indicators for the presence of each interaction type, set to 1 when present and 0 otherwise.
Because not every feature is computable for every task instance, we encode unavailable behavioral features using a sentinel value of -1.
Table~\ref{tab:behavioral_feature_table} lists our behavioral features.

\vspace{-1mm}
\subsection{Multi-Class Classification}
\label{sec:classifier}
To evaluate whether browser and behavioral fingerprints are sufficiently discriminative to distinguish among {\aibot}s and between {\aibot}s and humans, we formulate the problem as multi-class classification.
Due to a limited number of participants, we end up collecting 546 instances of humans data as compared to 1000 instances per {\aibot} across the three tasks, leading to moderate class imbalance.
We use XGBoost~\cite{xgboost2026xgboost} to train tree-based classifiers that mitigate class imbalance, handle heterogeneous features across different numerical scales, and learn non-linear interactions.
To quantify the contribution of each feature modality, we train classifiers on three feature sets: browser fingerprints, behavioral fingerprints, and their combination.
We also train classifiers on {\aibot}s alone before adding the human class, allowing us to compare feature importance and classification performance across class sets.
In total, this yields six classifiers, each evaluated using an 80--20 train-test split.

\vspace{-1mm}
\subsection{Analysis Methodology}

\para{Classifier analysis}
To analyze \texttt{FP-Agent}'s performance and interpret which features drive classification, we use both XGBoost feature importance and SHapley Additive exPlanations (SHAP) \cite{shap2026docs}.
For XGBoost, we use total gain as the feature-importance metric to quantify each feature's aggregate contribution to the classifier across all decision splits.
We then apply SHAP to obtain a finer-grained view of feature contributions by class.

\para{Statistical validation}
Many of the features we compare are continuous, non-normally distributed, and may exhibit unequal variance across classes.
Accordingly, we use the Mann-Whitney U test \cite{chicco2025simple} for pairwise comparisons and report $p$-values together with effect sizes $r \in [-1,1]$ computed using rank-biserial correlation, where the sign indicates the direction of the comparison.
We use Cohen's thresholds for Pearson's coefficient $r$ to quantify the effect sizes (large $\geq0.5$, medium $\geq 0.3$, small $\geq 0.1$) \cite{fiel2026effect}.
To compare feature variance between classes, we use the Brown-Forsythe test, which is robust to non-normality \cite{brown1974robust}, and report standard deviation ratios (SD ratios) to indicate directionality.
For both tests, we consider $p < 0.01$ statistically significant.

%% file: sections/05_characterizing_fingerprints.tex
\vspace{-1mm}
\section{Characterizing Fingerprints}
\label{sec:characterizing_fingerprints}
In this section, we show how and why browser and behavioral fingerprints distinguish among {\aibot}s and between {\aibot}s and humans.
We first assess classifier performance in Section \ref{sec:classifier_evaluation} to quantify the discriminative power of each feature set.
We then characterize browser and behavioral fingerprints in Sections \ref{sec:fpjs_analysis} and \ref{sec:behavioral_fp_analysis}, respectively, to explain the observed classification performance.

\begin{table*}
    \centering
    \setlength{\tabcolsep}{4.5pt}
    \caption{Overview of all classes and their browser fingerprints. ``*'' denotes paying user base size; ``-'' means unknown. ``Unique FPs'': number of unique fingerprints observed. ``Shared FPs'': fingerprints also produced by other classes.
    }
    \label{tab:fpjs_table}
    \vspace{-2.5mm}
    \input{tables/unique_browser_fingerprints}
    \vspace{-2.5mm}
\end{table*}

\vspace{-1mm}
\subsection{Classifier Evaluation}
\label{sec:classifier_evaluation}

We evaluate the six classifiers spanning three feature and two class sets as described in section~\ref{sec:classifier}.
We evaluate classifier performance using precision, recall, and $F_1$ score, reported in Table~\ref{tab:classifier_variant_eval}.
We further examine per-class performance using their confusion matrices in Figure~\ref{fig:confusion_matrices}.

\para{Performance}
The behavioral- and combined-fingerprint classifiers achieve near-perfect $F_1$ scores, which is also reflected in their confusion matrices showing almost no misclassifications.
In contrast, the browser-fingerprint classifier yields an $F_1$ score of approximately 0.8, indicating that browser fingerprints provide weaker discriminative power than behavioral fingerprints.
The slight increase in $F_1$ score when adding the human class further suggests that {\aibot}s and humans exhibit characteristic differences in the fingerprint features we collect.

\para{Misclassifications}
The relatively lower performance of the browser-fingerprint classifier can be explained by its confusion matrix, which reveals its susceptibility to false positives and false negatives, particularly when distinguishing amongst Atlas Agent, Browser Use, and Claude.
To understand the root cause of such misclassifications, we examine different browser fingerprints that each {\aibot} produces.
Table~\ref{tab:fpjs_table} highlights the number of unique fingerprints for each class, their distributions, and which classes share fingerprints with other classes.
We find that only Atlas Agent, Browser Use, and Claude share fingerprints.
When multiple classes produce the same fingerprint, the classifier assigns that fingerprint to the class with which it is most frequently associated in training.
For example, all three {\aibot}s produce the same browser fingerprint $f$ on macOS.
Because Atlas Agent is tested exclusively on macOS, all Atlas Agent trials produce $f$, whereas only a subset of Browser Use and Claude trials do.
As a result, the classifier systematically labels $f$ as Atlas Agent, explaining why some Browser Use and Claude instances are misclassified as Atlas Agent in Figure~\ref{fig:confusion_matrices}.
Our human participants performed tasks on their own sys- 

\noindent tems, which reduces overlap between human and {\aibot} browser fingerprints in our dataset.
Accordingly, the absence of shared browser fingerprints between humans and {\aibot}s in our data should not be interpreted as evidence that such overlap would not occur in practice.

\begin{tcolorbox}[takeawaybox]
     \textbf{Key Takeaway:} Browser fingerprints alone are insufficient to reliably distinguish among {\aibot}s and humans, particularly when multiple classes share the same browser fingerprint.
\end{tcolorbox}

\vspace{-2mm}
\subsection{Characterizing Browser Fingerprints}
\label{sec:fpjs_analysis}
Next, we analyze the browser fingerprints of {\aibot}s and human participants in our study, as shown in Table~\ref{tab:fpjs_table}.

\subsubsection{\textbf{Correlated browser fingerprints}}
We observe that, with the exception of Skyvern, local {\aibot}s produce stable browser fingerprints within a given test environment.
For example, Atlas Agent, Browser Use, and Claude consistently report the same screen resolution across trials on the same underlying system: 1440x900 on macOS and 1536x864 on Windows.
More generally, the number of unique browser fingerprints observed for each local {\aibot} closely tracks the number of systems on which it was tested, indicating that these fingerprints largely reflect the test environments rather than active fingerprint variation.
We discuss implications of this correlation in Section~\ref{sec:generalizability}.
We observe a similar pattern for cloud-based {\aibot}s.
For example, browser fingerprints for Manus consistently report Linux x86-64 as the system, suggesting that Manus runs on the same or highly similar remote environments across trials.

\begin{tcolorbox}[takeawaybox]
    \textbf{Key Takeaway:} Browser fingerprints of {\aibot}s are  correlated with their underlying execution environment, suggesting that current implementations do not actively vary these fingerprints.
\end{tcolorbox}

\begin{table}[b]
    \centering
    \small
    \caption{Classifier metrics across feature sets.}
    \label{tab:classifier_variant_eval}
    \vspace{-2mm}
    \input{tables/classifier_variant_eval}
\end{table}

\subsubsection{\textbf{Unique browser fingerprints}}
Having examined shared browser fingerprints among Atlas Agent, Browser Use, and Claude, we now analyze the unique fingerprint attributes of ChatGPT Agent, Manus, Comet, and Skyvern---the {\aibot}s that do not share browser fingerprints with other classes.

\para{ChatGPT Agent}
ChatGPT Agent uses a Chromium-based browser on a system with a screen resolution of 1280x960 and no HDR support throughout our experiments.
The reported system alternates between Linux x86\_64 and MacIntel, suggesting that it runs on at least two distinct remote environments.
Unlike the other classes, ChatGPT Agent consistently reports a font list containing only Calibri and 13 CPU cores irrespective of the reported system.
We also observe variation in its plugins list, which is either empty or populated with the default PDF plugins typically found in web browsers.
In addition, ChatGPT Agent produces significantly larger font preference measurements than all other classes ($p < 0.01$, $r = 1.00$).
Because font preferences reflect the number of pixels occupied by rendered text, these unusually large values may indicate that text is rendered at an enlarged size for improved visual readability.

\para{Manus}
Manus performs 96.6\% of its tasks on a Linux x86\_64 system with no HDR support and using a Chromium-based browser.
It also consistently reports 4GB of RAM and 6 CPU cores, suggesting the use of a relatively lightweight execu- 

\noindent tion environment.
Other notable attributes include a screen resolution of 1280x1100, UTC timezone, no installed fonts, and a max touch points value of 10.

Despite all local {\aibot}s using Chromium-based browsers and, where applicable, the same underlying systems, Comet and Skyvern do not share browser fingerprints with Atlas Agent, Claude, Browser Use, or each other.
We therefore compare browser fingerprints among locally tested {\aibot}s on the same supported systems to identify the distinguishing attributes of Comet and Skyvern.

\para{Comet}
On both Windows and macOS, Comet's browser fingerprint is largely similar to those of the other local {\aibot}s.
On macOS, the main difference is that Comet exhibits significantly larger font preference values ($p < 0.01$, $r = 1.00$) than Atlas Agent, Browser Use, and Claude, which may likewise reflect enlarged text rendering.
On Windows, however, its font preference values are consistent with Claude and Browser Use.
Instead, the distinguishing attribute on Windows is that Comet reports the Gill Sans font, whereas Claude and Browser Use report Gill Sans only on macOS.

\para{Skyvern}
Skyvern is notable in that it appears to override parts of its browser fingerprint.
On macOS, Claude, Atlas Agent, Comet, and Browser Use all report the underlying system's attributes---for example, a system value of MacIntel, screen resolution of 1440x900, and 8 CPU cores.
In contrast, Skyvern preserves some attributes, such as font preferences and system, while modifying others, including timezone and screen resolution.
Table~\ref{tab:skyvern_fpjs_differences} lists the Skyvern-reported attributes that differ from the other local {\aibot}s on macOS.
On Linux, Skyvern reports the same modified values as on macOS, indicating that these attributes are likely configured independently of the underlying system.

\begin{tcolorbox}[takeawaybox]
    \textbf{Key Takeaway:} {\capitalizeaibot}s that do not share browser fingerprints differ through distinctive fingerprint attributes, arising either from variation in their remote execution environments or from  customization of reported browser properties.
\end{tcolorbox}

\subsection{Characterizing Behavioral Fingerprints}
\label{sec:behavioral_fp_analysis}
Because browser fingerprints provide limited discriminative power by themselves, we next analyze behavioral fingerprints to explain the strong performance of the behavioral- and combined-fingerprint classifiers.
We focus on three common interaction modalities (typing, scrolling, and mouse movement) that capture how {\aibot}s and humans interact with websites.

\subsubsection{\textbf{Typing behavior.}}
\label{sec:typing_behavior}
\begin{figure}
    \centering
    \includegraphics[width=0.85\linewidth]{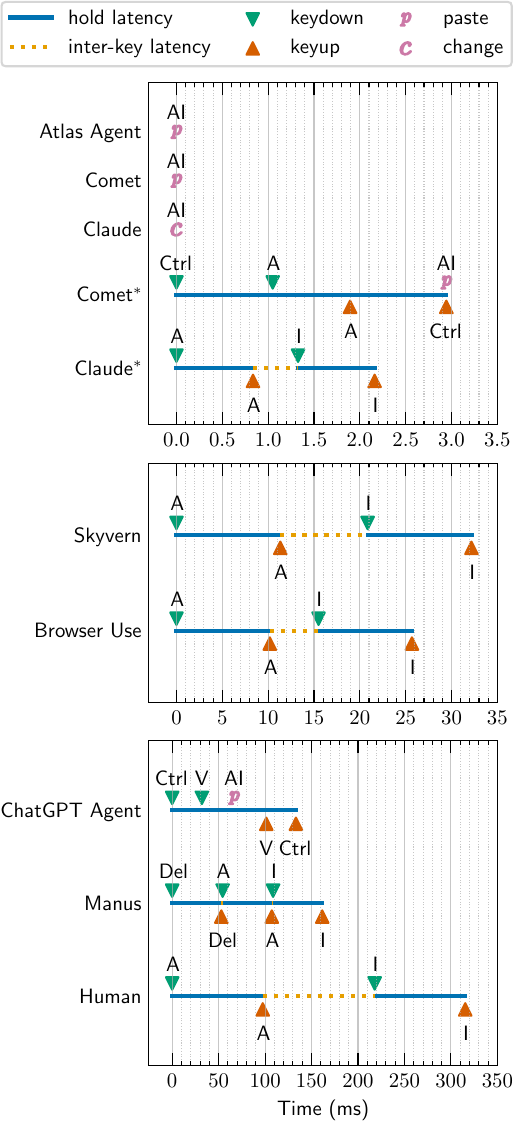}
    \vspace{-2.5mm}
    \caption{Time series of events representing each {\aibot} and human participants' averaged typing behavior. The scenario assumes entering ``AI'' when the text field is in. ``*'' indicates non-default behavior.}
    \Description[]{}
    \label{fig:typing_dynamics}
    \vspace{-2mm}
\end{figure}

Typing-related features are among the most informative behavioral features in \texttt{FP-Agent}.
These include inter-key latency, hold latency, backspace/delete usage, and the number of DOM input and change events.
We define a keystroke as a sequence of a \texttt{keydown} event followed by a \texttt{keyup} event.
Figure~\ref{fig:typing_dynamics} visualizes the average typing behavior of each {\aibot} and our human participants.
The mean, 95\% confidence intervals, and standard deviation of inter-key and hold latencies for each class can be found in Table \ref{tab:keystroke_stats} in Appendix~\ref{appendix:keystroke_latency_stats}.

\begin{table}[t]
    \centering
    \setlength{\tabcolsep}{4.5pt}
    \small
    \caption{Skyvern browser fingerprint attributes as compared to other {\aibot}s on MacOS. ``Other'' includes Atlas Agent, Browser Use, Claude, and Comet.}
    \label{tab:skyvern_fpjs_differences}
    \vspace{-2.5mm}
    \input{tables/skyvern_fpjs_differences}
    \vspace{-4mm}
\end{table}

\para{Paste-based input}
Our data shows that ChatGPT Agent, Atlas Agent, and Comet paste text into textual input fields rather than typing character-by-character.
Among these three {\aibot}s, ChatGPT Agent is the only one that pastes using the Ctrl+V keyboard shortcut.
While entering Ctrl+V, ChatGPT Agent produces high hold latencies ($p < 0.01$, $r = 1.00$), with an average of 66.58ms and a standard deviation of 34.46ms.
The variance occurs because Ctrl is pressed before and released after V, giving the Ctrl keystroke a longer hold latency than V.
We also observe an inconsistency between the reported system and the keyboard shortcut used by ChatGPT Agent: when reporting MacIntel as its system, it still uses Ctrl+V rather than the expected macOS shortcut Cmd+V.

\para{Paste event input}
On macOS, Atlas Agent and Comet both appear to type by programmatically firing a paste event.
On Windows, Comet additionally uses the Ctrl+A shortcut to select the entire text field before firing the paste event, likely to overwrite any preexisting text.
When producing this keyboard shortcut, Comet exhibits a low average hold latency of 2.97ms with a standard deviation of 1.66ms.
As with ChatGPT Agent, this variance arises from the temporal overlap between modifier and non-modifier keys, though Comet's hold latencies are much lower ($p < 0.01$, $r = 1.00$).

\para{Keystroke input}
Browser Use, Skyvern, and Manus type using keystrokes, each with distinct inter-key and hold latencies.
We observe average inter-key latencies of 5.31ms, 9.52ms, and 1.39ms, and average hold latencies of 10.19ms, 11.33ms, and 52.92ms for Browser Use, Skyvern, and Manus, respectively.
Like ChatGPT Agent, Manus exhibits high hold latencies ($p < 0.01$, $r = 1.00$) with little variance.

We additionally examine the source code of Skyvern and Browser Use~\cite{skyvern2026skyvernrepo, browseruse2026browseruserepo} to assess whether their typing implementations align with our measurements.
Skyvern uses Playwright for browser automation, whereas Browser Use relies on a custom interface~\cite{cdpuse2026cdpuserepo} to the Chrome DevTools Protocol (CDP).
With $n$ denoting the length of the input text, Skyvern programmatically inputs the first $n - 20$ characters when $n > 20$, and uses keystrokes with a 10ms inter-key latency for the final 20 characters~\cite{skyvern2026handlerutils}.
This behavior is corroborated by our measurements.
We do not find any explicit hold-latency setting in Skyvern's implementation.
We discuss Skyvern's programmatic inputs in paragraph \textbf{Input events.}

Browser Use types with the \texttt{Input.dispatchKeyEvent} 

\noindent CDP method~\cite{google2026dispatchkeyevent}, and uses Python's \texttt{asyncio.sleep()}~\cite{python2026asynciosleep} to program an inter-key latency of 18ms and a hold latency of 1ms~\cite{browseruse2026element}.
However, we observe an average inter-key latency of 5.31ms and an average hold latency of 10.19ms (Figure~\ref{fig:typing_dynamics}).
This discrepancy likely arises from event buffering due to renderer scheduling latency~\cite{harrelson2021renderingng}, causing buffered events to be executed in bursts.
As a result, events spaced 18ms apart in the automation logic may be processed closer together by the renderer, reducing the observed inter-key latency.
The same renderer-side delay likely explains why the observed hold latency exceeds the observed inter-key latency, as processing and rendering text changes occurs during the hold duration.

\para{Change event input}
By default, Claude programmatically fills text fields using only a change event.
However, we find that Claude is also capable of keystroke-based typing.
In the forums task, we explicitly instruct {\aibot}s to type using keystrokes and avoid pasting.
Claude is the only {\aibot} that switches from programmatic to keystroke input in response.
When it does so, it produces almost no inter-key or hold latency, with both below 1ms.

\begin{tcolorbox}[takeawaybox]
    \textbf{Key Takeaway:} {\capitalizeaibot}s exhibit highly heterogeneous typing behavior, ranging from programmatic filling of text fields (Atlas Agent, ChatGPT Agent, Claude, Comet) to keystroke-based input with latencies in tens of milliseconds (Manus), to few milliseconds (Skyvern and Browser Use), and to tenths of a millisecond (Claude). 
\end{tcolorbox}

\para{Human typing}
Unsurprisingly, human typing differs substantially from {\aibot} typing.
Across our human-participant sessions, we observe not only much higher inter-key ($p < 0.01$, $r = 1.00$) and hold ($p < 0.01$, $r = 0.98$) latencies, but also substantially greater variance in inter-key ($p < 0.01$, $r = 1.00$) and hold latencies ($p < 0.01$, $r = 0.61$) than any {\aibot}.
Accordingly, both the mean and the standard deviation of inter-key and hold latency serve as strong signals for distinguishing humans from {\aibot}s.
We also frequently observe humans beginning the next keystroke before releasing the current one, producing a down-down-up-up rather than down-up-down-up event sequence.
Although our features do not explicitly encode this overlap, they are still sufficient to distinguish human typing from {\aibot} typing.

\begin{tcolorbox}[takeawaybox]
    \textbf{Key Takeaway:} Humans exhibit much higher inter-key and hold latencies, together with substantially greater variance, than {\aibot}s.
\end{tcolorbox}

\para{Backspace/delete keystrokes}
Unlike humans, {\aibot}s are not expected to make spontaneous typing mistakes in the same way that humans may do.
We therefore examine the total number of backspace and delete keystrokes, as well as the fraction of keystrokes corresponding to these keys.
Among the {\aibot}s that produce keystrokes, only Manus and Skyvern press delete, with delete accounting for 7\% and 2\% of keystrokes on average, respectively.
For Skyvern, we observe repeated cycles in which semantically equivalent text is typed and then fully deleted before the next action.
This yields a number of delete keystrokes roughly proportional to the number of text fields.
One possible explanation is Skyvern trying multiple LLM-generated input variants for a field before settling on one.
We do not observe Manus deleting text after typing during the task, so we manually inspect its event logs to determine when delete is pressed.
We find that Manus presses delete before beginning to fill out a text field, which may reflect a field-clearing behavior similar to Comet on Windows.
Human participants produce backspace/delete keystrokes at rates comparable to Manus and Skyvern.
However, for humans these keystrokes plausibly arise from ordinary typing mistakes, revisions, or reformulations, suggesting that backspace/delete events may still be informative for distinguishing humans from agents.

\begin{figure*}
    \centering
    \includegraphics[width=0.9\linewidth]{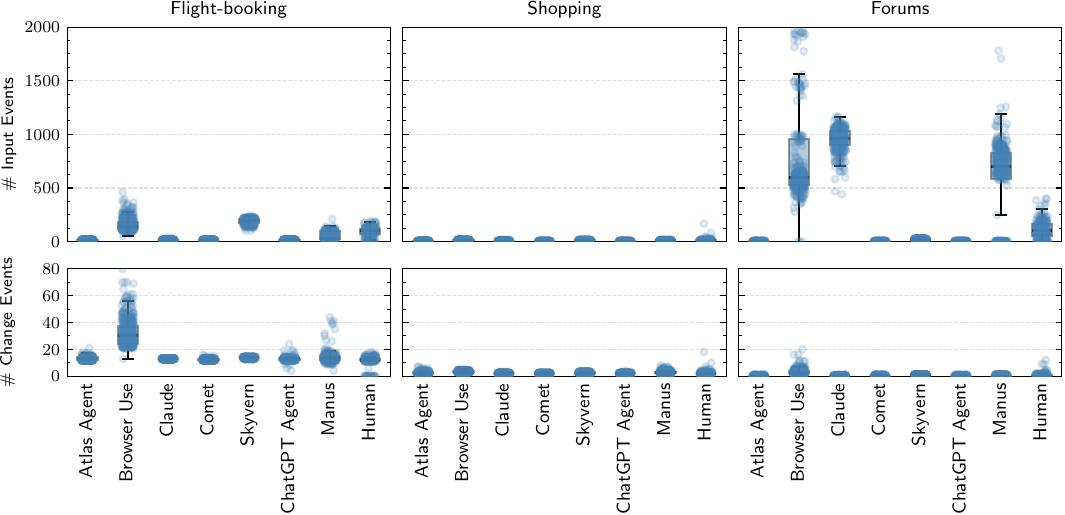}
    \vspace{-2mm}
    \caption{Strip plot of number of change and input events by task for each {\aibot} and humans.}
   \Description[]{Strip plot over a box plot of {\aibot}s' number of change and input events by task.}
   \label{fig:change_input_events_plots}
   \vspace{-2mm}
\end{figure*}

\para{Change events}
Our SHAP analysis identifies the number of change events as an informative feature for distinguishing Browser Use from the other classes.
A change event occurs when the value of an \texttt{input}, \texttt{select}, or \texttt{textarea} element changes~\cite{mdn2025changeevent}.
For text fields, change events fire only when the element loses focus, reducing overlap with input events, which fire with each character entry.
We visualize change-event counts in Figure~\ref{fig:change_input_events_plots}.
Browser Use produces more change events than the other {\aibot}s across all tasks on average ($p < 0.01$, $r = 0.454$).
In the flight-booking task, which contains many text form fields, Browser Use produces roughly twice as many change events as the other {\aibot}s.
This pattern is consistent with its source code, which clears text fields using an extra input and change event before entering text.
In contrast, human participants are largely indistinguishable from {\aibot}s on this feature because the number of elements that must be changed is largely determined by the task rather than by who performs it.

\begin{tcolorbox}[takeawaybox]
    \textbf{Key Takeaway:} Browser Use is distinctive in producing double the change events than the other {\aibot}s when filling text fields.
\end{tcolorbox}

\para{Input events}
Our feature-importance analysis identifies input-event counts as another informative feature.
To compare input-event counts across classes, we plot counts by task in Figure~\ref{fig:change_input_events_plots}.
An input event fires whenever the value of an input element changes~\cite{mdn2025inputevent}.
In text fields, this means that an input event is typically generated for each entered character, making it correlated with keystroke-based typing behavior.
Our data shows that classes that type with keystrokes (Browser Use, Manus, and humans) have a higher average number of input events ($p < 0.01$, $r = 0.431$) and greater variance ($p < 0.01$, SD ratio: 1.90) than classes that rely on paste or programmatic input.
The two exceptions are Claude and Skyvern.
For Claude, Figure~\ref{fig:change_input_events_plots} shows two input-event distributions corresponding to its two typing modes: change-event-based input in the flight-booking and shopping tasks, and keystroke-based input in the forums task when explicitly prompted to type.
For Skyvern, the distribution largely disappears in the forums task because that task requires a long text response in a single field, and Skyvern types at most 20 characters using keystrokes before switching to programmatic input.
This behavior matches the implementation discussed earlier (\textbf{Keystroke input}).
As with change events, human input-event counts remain highly variable and task-dependent.
For example, humans may produce input-event counts similar to those of keystroke-based {\aibot}s when entering the same text, even with mistakes.
This helps explain why input-event counts are more useful as a supplement to other typing features than as a primary feature for distinguishing humans from {\aibot}s.

\begin{tcolorbox}[takeawaybox]
    \textbf{Key Takeaway:} Input-event counts complement other typing features by distinguishing keystroke-based input from paste- and programmatic-input strategies.
\end{tcolorbox}

\begin{figure*}
    \centering
    \includegraphics[width=0.9\linewidth]{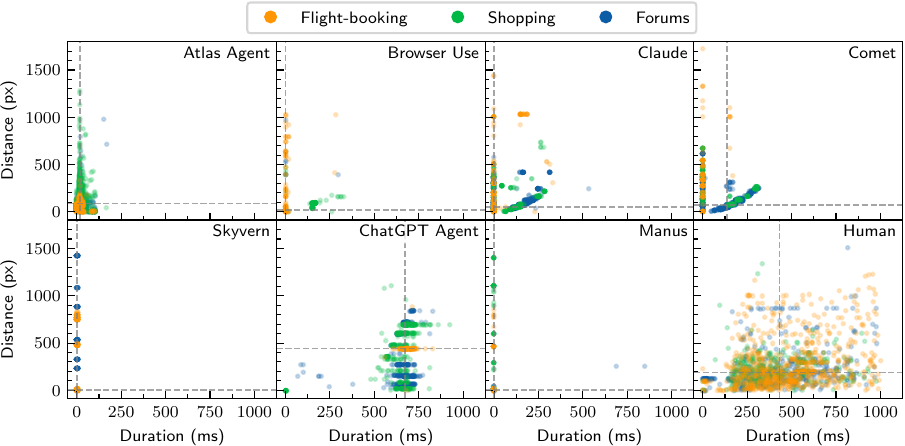}
    \vspace{-2mm}
    \caption{Scroll distance vs duration for each class. Each point represents a scroll burst. The dotted lines are at the class median for their respective axis.}
    \Description[]{Scroll distance vs duration scatter plots for each class.}
    \label{fig:scroll_distributions}
    \vspace{-2.5mm}
\end{figure*}

\subsubsection{\textbf{Scroll behavior}}
Scroll-related features also rank consistently high in both overall feature-importance analysis and per-class SHAP values.
To visualize differences in scroll behavior across classes, we plot scroll distance versus duration by task in Figure~\ref{fig:scroll_distributions}.
Because many points overlap in the plots, we use higher opacity to indicate denser regions.
Across all experiments, {\aibot}s exhibit non-continuous scrolling, in which the page position changes in discrete jumps rather than through smooth, continuous movement.
Based on these plots, we observe two notable patterns of such non-continuous scrolling, attributable to differences in browser agent architectures as discussed in Section~\ref{sec:ai_bots}.

\para{Instant scrolling of an element into view}
One way {\aibot}s scroll is by instantaneously scrolling an element into view.
We believe this behavior occurs when the {\aibot} has access to the webpage's DOM as context and knows exactly where to scroll to.
Skyvern and Manus exhibit this pattern most clearly.
In addition, Skyvern, Browser Use, and Manus all show clusters of points at specific scroll distances with duration 0ms, indicating that they repeatedly scroll to similar positions for a given task.
Because Skyvern uses Playwright's default element-scrolling behavior, which can instantly scroll elements into view, these clusters are likely explained by its underlying automation implementation.
Since Manus and Browser Use exhibit similar patterns, they may also use a similar scrolling implementation.

\para{Multiple short, non-continuous scroll bursts}
A second pattern consists of sequences of shorter, discrete scroll bursts rather than a single direct jump.
This behavior is consistent with incremental exploration of the page, where the {\aibot} scrolls in stages and evaluates intermediate views before proceeding.
In our data, these bursts appear either as short zero-duration jumps without strong clustering or as longer bursts with nonzero duration.
Atlas Agent, Browser Use, Claude, Comet, and ChatGPT Agent all exhibit this pattern.
ChatGPT Agent shows larger average scroll distance and duration than the other {\aibot}s in this group ($p < 0.01$, $r \approx 0.6$), representing a more extreme version of the same behavior.
Taken together, these observations suggest that some {\aibot}s rely more heavily on incremental viewport-based interaction, whereas others more often scroll directly to recurring target positions.
Browser Use likely combines both styles, as it exhibits both clustered instantaneous jumps and multi-burst scrolling behavior.

\begin{tcolorbox}[takeawaybox]
    \textbf{Key Takeaway:} {\capitalizeaibot}s implement scrolling through either direct jumps, suggesting DOM-based task execution, or multiple scroll bursts, suggesting vision-based task execution.
\end{tcolorbox}

\para{Human scrolling}
Human participants exhibit significantly greater scroll duration ($p < 0.01$, $r = 0.67$) and distance ($p < 0.01$, $r = 0.39$) than most {\aibot}s.
The sole exception is ChatGPT Agent, which exhibits higher average scroll duration and distance ($p < 0.01$, $r = 0.35$) than our participants.
Participants also exhibit higher variance in scroll duration than all {\aibot}s, and higher variance in scroll distance than all {\aibot}s except Manus, Skyvern, and ChatGPT Agent ($p < 0.01$ and SD ratio $> 1$ for all pairwise comparisons).
Thus, summary statistics of scroll distance and duration already provide substantial signal for distinguishing humans from {\aibot}s, while the overall distribution of scroll behavior provides additional information that can strengthen this distinction.

\begin{tcolorbox}[takeawaybox]
    \textbf{Key Takeaway:} Humans generally exhibit longer and more variable scrolling than most {\aibot}s. The main exceptions are Manus, Skyvern, and ChatGPT Agent, which show comparatively high variance in scroll distance, while ChatGPT Agent also exceeds humans in average scroll distance and duration.
\end{tcolorbox}

\subsubsection{\textbf{Mouse behavior}}
We define mouse movement as a sequence of consecutive \texttt{mousemove} events terminated by either a click (\texttt{mousedown} and \texttt{mouseup}) or an idle period.
We set the idle threshold to 250ms, which lies within the range of human reaction times \cite{bionumbers2026reactiontime, humanbenchmark2026reactiontimetest}.
Across our experiments, {\aibot}s produce mouse events only at click time, yielding an event sequence consisting of a single \texttt{mousemove} followed immediately by \texttt{mousedown} and \texttt{mouseup} at the click location.
Thus, we do not observe continuous pointer movement between click locations for {\aibot}s; instead, the pointer appears to jump directly to the target location.
In contrast, humans generate continuous streams of mouse-move events, allowing us to compute features such as curvature angle, curvature distance, and movement direction.
As a result, mouse-movement features are among the strongest signals for distinguishing humans from {\aibot}s, as reflected in the high importance of mouse movement angle-of-curvature range in Table~\ref{tab:behavioral_feature_table}.

\begin{tcolorbox}[takeawaybox]
    \textbf{Key Takeaway:} {\capitalizeaibot}s do not exhibit any mouse movements and instead directly teleport to the click location, making them  distinguishable from humans.
\end{tcolorbox}

%% file: tables/unique_browser_fingerprints.tex
\begin{tabular}{lccccccccc}
    \toprule
    \multirow{2}{*}{Class} & User base & Deployment & Open & Systems & Unique & Top-1 & Norm. & Shared & Shared \\
    & Size & Model & Source? & Tested & FPs & Coverage & Entropy & FPs & With \\
    \midrule
    Atlas Agent \ $\vcenter{\hbox{\atlasicon}}$ & 50M$^*$~\cite{openai2026scalingai} & Local & No & $\vcenter{\hbox{$\vcenter{\hbox{\macosicon}}$}}$ & 1 & 100.00\% & 0.00 & 1 & $\vcenter{\hbox{\browseruseicon}}$ \ $\vcenter{\hbox{\claudeicon}}$ \\ [2pt]
    Browser Use \ $\vcenter{\hbox{\browseruseicon}}$ & 89.1K $\vcenter{\hbox{\githubstaricon}}$~\cite{browseruse2026browseruserepo} & Local & Yes & $\vcenter{\hbox{\windowsicon}}$ \ $\vcenter{\hbox{\macosicon}}$ \ $\vcenter{\hbox{\linuxicon}}$ & 4 & 41.44\% & 0.79 & 2 & $\vcenter{\hbox{\atlasicon}}$ \ $\vcenter{\hbox{\claudeicon}}$ \\ [2pt]
    Claude \ $\vcenter{\hbox{\claudeicon}}$ & 6M~\cite{chrome2026claudeextension} & Local & No & $\vcenter{\hbox{\windowsicon}}$ \ $\vcenter{\hbox{\macosicon}}$ & 2 & 50.00\% & 1.00 & 2 & $\vcenter{\hbox{\browseruseicon}}$ \ $\vcenter{\hbox{\atlasicon}}$ \\ [2pt]
    Comet \ $\vcenter{\hbox{\cometicon}}$ & 2.4M$^*$~\cite{elad2025perplexityaistatistics} & Local & No & $\vcenter{\hbox{\windowsicon}}$ \ $\vcenter{\hbox{\macosicon}}$ & 2 & 51.17\% & 1.00 & 0 & N/A \\ [2pt]
    Skyvern \ $\vcenter{\hbox{\skyvernicon}}$ & 21.3K $\vcenter{\hbox{\githubstaricon}}$~\cite{skyvern2026skyvernrepo} & Local & Yes & $\vcenter{\hbox{\macosicon}}$ \ $\vcenter{\hbox{\linuxicon}}$ & 2 & 51.31\% & 1.00 & 0 & N/A \\ [2pt]
    ChatGPT Agent \ $\vcenter{\hbox{\chatgpticon}}$ & 50M$^*$~\cite{openai2026scalingai} & Cloud & No & N/A & 7 & 64.32\% & 0.51 & 0 & N/A \\ [2pt]
    Manus \ $\vcenter{\hbox{\manusicon}}$ & - & Cloud & No & N/A & 5 & 96.56\% & 0.11 & 0 & N/A \\ [2pt]
    Human \ $\vcenter{\hbox{\humanicon}}$ & N/A & N/A & N/A & N/A & 57 & 12.50\% & 0.95 & 0 & N/A \\
    \bottomrule
\end{tabular}

%% file: tables/classifier_variant_eval.tex
\begin{tabular}{ccccc}
    \toprule
    Class Set & Feature Set & Precision & Recall & $F_1$ \\
    \midrule
    \multirow{3}{*}{{\uppercaseaibot}s} & Browser & 0.8674 & 0.8179 & 0.7969 \\
    & Behavioral & 0.9993 & 0.9994 & 0.9993 \\
    & Combined & 1.0000 & 1.0000 & 1.0000 \\
    \midrule
    {\uppercaseaibot}s & Browser & 0.8840 & 0.8407 & 0.8223 \\
    + & Behavioral & 0.9994 & 0.9994 & 0.9994 \\
    Human & Combined & 1.0000 & 1.0000 & 1.0000 \\
    \bottomrule
\end{tabular}

%% file: tables/skyvern_fpjs_differences.tex
\begin{tabular}{ccc}
    \toprule
    Attribute & Skyvern & Other \\
    \midrule
     Screen Resolution & 1920x1080 & 1440x900 \\
     HDR & No & Yes \\
     Color Gamut & srgb & p3 \\
     Plugins & None & Default PDF plugins \\
     Timezone & America/New York & America/Los Angeles \\
     \bottomrule
\end{tabular}

%% file: sections/06_discussion_and_limitations.tex
\section{Discussion, Limitations, and Implications}
\label{sec:discussion}
In this section, we discuss the broader implications, scope, and durability of our findings.
We first examine the generalizability of our results in Section~\ref{sec:generalizability}, then consider how detection may evolve as part of an arms race in Section~\ref{sec:arms_race}.
Finally, we discuss deployment-oriented implications through real-time classification in Section~\ref{sec:realtimeclassification} and the effectiveness of industrial solutions in Section~\ref{sec:effectiveness_of_industrial_solutions}.

\subsection{Generalizability}
\label{sec:generalizability}
We assess how well our findings may extend beyond the specific agents, systems, participants, and tasks included in our study.
We first discuss the main sources of limited generalizability in our measurement design, then evaluate generalization to unseen tasks by simulating a held-out task.

\begin{table}
    \centering
    \small
    \caption{Classifier performance for the task generalization experiment. ``*'' denotes the task used as test set.}
    \label{tab:generalization_experiments_eval}
    \vspace{-2.5mm}
    \input{tables/generalization_experiments_eval}
    \vspace{-7mm}
\end{table}

\para{Coverage of browsing agents}
Our study measures the characteristics of seven {\aibot}s and evaluates classification in a closed-world setting, where each class corresponds to one of the studied {\aibot}s or to humans.
Accordingly, \texttt{FP-Agent} is not designed to identify unseen {\aibot}s as distinct agent classes.
However, our FP-Agent framework itself is readily extensible: new {\aibot}s can be incorporated by collecting artifacts for them and adding them as new classes during training.
Moreover, the strong separation between {\aibot}s and humans in our behavioral analyses suggests that some behavioral distinctions may potentially generalize beyond the specific browsing agents we study in this paper.

\para{Coverage of system configurations}
Although we test each {\aibot} on each system they support, the space of possible hardware, operating system, browser, and display configurations is much larger than we can exhaustively cover.
As a result, our study captures only a subset of the browser fingerprints that may arise in practice.
Expanding coverage to a broader range of system configurations would likely improve robustness to environment-specific differences and generalizability of browser-fingerprint-based classification.
The \texttt{FP-Agent} framework is also extensible along this dimension.
At the same time, our measurements already capture meaningful overlaps and distinctions across browser finger-

\noindent prints, including cases where multiple {\aibot}s share the same underlying browser fingerprint and cases where similar environments still yield distinctive fingerprints.

\para{Coverage of human behavior}
Our study includes a substantial human dataset, but it is drawn from a relatively narrow population of internet users: undergraduate students.
Accordingly, we acknowledge that the human browser and behavioral fingerprints in our dataset do not capture the full diversity of human behavior that may arise in practice.
Our goal, however, is not to model the entire space of human interaction, but to determine whether humans can be distinguished from {\aibot}s under realistic task execution.
Within that scope, our human data provides sufficient evidence of difference between humans and {\aibot}s.

\para{Coverage of interaction tasks}
Our three tasks do not capture the full diversity of website layouts and interaction patterns that {\aibot}s may encounter in practice.
Instead, they are designed to cover common web interaction primitives under controlled conditions.
As a result, less common interaction patterns may be absent from our dataset.
For example, we do not include drag-and-drop interactions because not every {\aibot} we study supports them.
To assess how sensitive our results are to task choice, we perform an additional held-out-task generalization experiment.
We split the data by task into a training set $R$ containing data from two of the three tasks and a test set $E$ containing data from the remaining task.
Because browser fingerprints are task-independent, we do not evaluate a browser-fingerprint classifier in this experiment.
Instead, we only train behavioral- and combined-fingerprint classifiers on $R$.
We repeat this procedure for all choices of held-out task.

The results are shown in Table~\ref{tab:generalization_experiments_eval}.
The performance of behavioral-fingerprint classifier drops substantially on unseen tasks, with $F_1$ decreasing by as much as 0.369 relative to Table~\ref{tab:classifier_variant_eval}.
The larger drop in behavioral-fingerprint classifier performance is expected because our tasks intentionally exercise partially non-overlapping interaction patterns.
Even when these interactions generate the same event types, the resulting behavioral features can differ meaningfully across tasks.
Consequently, when the classifier encounters interaction patterns during testing that were absent during training, its performance degrades.
At the same time, the three tasks collectively span the most common web interaction primitives, suggesting that our results capture a meaningful portion of real-world browsing behavior even if they do not fully characterize it.
More broadly, the \texttt{FP-Agent} framework is extensible: incorporating additional tasks would expand coverage of interaction patterns and likely improve robustness to unseen-task scenarios.
Notably, the combined-fingerprint classifier degrades negligibly, with a maximum $F_1$ drop of only 0.0684.
This result shows that although browser fingerprints alone are insufficient, they still provide useful task-agnostic signal that improves robustness in unseen-task settings.

\vspace{-1mm}
\subsection{Arms Race}
\label{sec:arms_race}

\para{Evolving fingerprints}
Our results suggest that existing {\aibot}s do not actively evade detection out of the box.
In their present form, they remain distinguishable from humans and from each other using a combination of browser and behavioral fingerprint features.
However, these fingerprints should not be assumed to be stable over time.
For Skyvern and Browser Use, the open-source {\aibot}s in our study, we fix the evaluated versions to ensure reproducibility.
For proprietary {\aibot}s, however, we cannot control updates during the measurement period.
Despite this, we do not observe substantial fingerprint drift for most proprietary {\aibot}s during our experiments.
The main exception is ChatGPT Agent, which produces multiple browser fingerprints over time, likely reflecting changes in its underlying implementation.
This suggests that while current {\aibot} fingerprints are sufficiently stable, future updates may alter them.

\para{Agent humanization}
{\capitalizeaibot} detection, like traditional bot detection, is likely to result in an arms race.
Our results show that current {\aibot}s do not yet closely mimic human browser and behavioral fingerprints: their browser fingerprints often reflect stable execution environments, and their typing, scrolling, and mouse behaviors remain readily distinguishable from those of humans.
However, this gap may narrow as {\aibot}s become more sophisticated.
{\capitalizeaibot}s are \textit{capable} of producing human-like browser fingerprints and, leveraging LLM-based reasoning, can navigate websites in increasingly human-like ways, including solving CAPTCHA challenges.
As future {\aibot}s incorporate stronger behavioral imitation, the features identified in our work may weaken, making detection more challenging.
An important direction for future work is therefore to study more nuanced behavioral fingerprint features that better capture residual differences between humans and increasingly human-like {\aibot}s.

\vspace{-1mm}
\subsection{Real-time Classification}
\label{sec:realtimeclassification}

Real-time detection of {\aibot}s would allow websites to respond during an active session, for example by blocking the visitor, serving a different website version, or applying browsing-agent-specific controls.
At the same time, such deployment requires making accurate predictions from only a limited amount of behavioral data early in the interaction.

\begin{figure}
    \centering
    \includegraphics[width=0.85\linewidth]{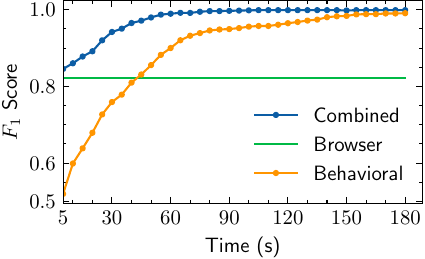}
    \vspace{-2.5mm}
    \caption{$F_1$ score over time elapsed into the task browsing session for each feature set.}
    \Description[]{$F_1$ score over time elapsed into the task browsing session for each feature set.}
    \label{fig:real-time_f1_score_vs_time}
    \vspace{-3mm}
\end{figure}

To evaluate whether our approach remains effective under such constraints, we measure \texttt{FP-Agent}'s performance over progressively larger observation windows, restricting behavioral data to incremental time windows of $\{5s, 10s, \dots, 180s\}$, where each window begins at the first observed behavioral event.
Figure~\ref{fig:real-time_f1_score_vs_time} reports the resulting $F_1$ scores over time.
Because browser fingerprints are static within a session, the browser-fingerprint classifier yields the same $F_1$ score across all time windows.
In contrast, the behavioral- and combined-fingerprint classifiers improve as more behavioral data becomes available, with performance plateauing after roughly 3 minutes and 1 minute, respectively.
These results suggest that, in our experimental setting, accurate real-time classification is feasible after relatively short observation periods, especially when browser and behavioral fingerprint features are combined.

\subsection{Effectiveness of Existing Defenses}
\label{sec:effectiveness_of_industrial_solutions}

\para{Limits of cryptographic self-identification}
As discussed in Section~\ref{sec:defenses_against_ai_bots}, many existing bot detection providers now incorporate \textit{Web Bot Auth} alongside their traditional detection methods~\cite{datadome2026agenttrust, human2026agentictrust, lee2025signedagents}, as cryptographic signatures provide stronger identity guarantees than user agent strings.
This is a useful step forward, but it does not eliminate the need for {\aibot} detection.
Web Bot Auth remains voluntary: {\aibot}s may choose not to participate (e.g., Atlas Agent~\cite{human2026atlas}, Browser Use, Comet~\cite{human2026comet}, Skyvern), and participating {\aibot}s may choose when to sign requests.
Accordingly, practical deployments still benefit from a fallback system like \texttt{FP-Agent} that can distinguish unsigned {\aibot} traffic from human traffic and attribute it to a specific class based on observed browser and behavioral fingerprints.

\begin{table}[!t]
    \centering
    \small
    \caption{Comparison of \texttt{FP-Agent} to Cloudflare's free bot management. ``*'' denotes blocking of the {\aibot} provider's non-task related server-side requests.}
    \label{tab:cloudflare_exp}
    \vspace{-2.5mm}
    \input{tables/cloudflare_experiment}
    \vspace{-4mm}
\end{table}

\para{Cloudflare case study}
As a concrete example of a widely deployed existing defense, we conduct a case study of Cloudflare.
We evaluate Cloudflare's free bot management \cite{cloudflare2026freebotplan}, which provides two AI-bot-related controls: \textit{AI Crawl Control}, which allows publishers to define rules to block \textit{specific} AI search bots, AI crawlers, and AI assistants; and \textit{Block AI Bots}, which provides a toggle to blanket-block \textit{all} AI crawlers~\cite{cloudflare2026bots}.
We enable both controls on our honey website and then prompt each {\aibot} to perform our three tasks.
The results are shown in Table~\ref{tab:cloudflare_exp}.
We find that only Manus is fully blocked by Cloudflare, whereas the other six {\aibot}s complete the tasks successfully.
Cloudflare does block requests from ClaudeBot and Perplexity-User, but IP-based analysis indicates that these requests originate from the {\aibot} providers' server-side infrastructure, not from the locally-running Claude and Comet themselves.
Cloudflare also blocks some requests from ChatGPT-User, yet ChatGPT Agent is still able to complete the tasks despite operating from cloud infrastructure.
Taken together, these results suggest that Cloudflare's detection is currently not fully effective at detecting {\aibot}s in our setting.
Moreover, Manus appears in Cloudflare's bot directory as a verified bot~\cite{cloudflare2026botdirectory}, suggesting that its successful blocking is likely based on self-identification.
More broadly, this highlights the need for defenses like \texttt{FP-Agent} that can work even when {\aibot}s do not voluntarily self-identify.

%% file: tables/generalization_experiments_eval.tex
\begin{tabular}{lcccc}
    \toprule
    Tasks & Features & Precision & Recall &$F_1$ \\
    \midrule
    \multirow{2}{*}{\makecell[l]{Shop, Flights,\\Forums$^*$}}
        & Behavioral & 0.8620 & 0.8350 & 0.7997 \\
        & Combined & 0.9496 & 0.9371 & 0.9316 \\
    \midrule
    \multirow{2}{*}{\makecell[l]{Shop, Forums,\\Flights$^*$}}
        & Behavioral & 0.7294 & 0.7021 & 0.6303 \\
        & Combined & 0.9718 & 0.9700 & 0.9681 \\
    \midrule
    \multirow{2}{*}{\makecell[l]{Flights, Forums,\\Shop$^*$}}
        & Behavioral & 0.7449 & 0.7838 & 0.7432 \\
        & Combined & 0.9741 & 0.9714 & 0.9708 \\
    \bottomrule
\end{tabular}

%% file: tables/cloudflare_experiment.tex
\newcommand{\tagyes}[1]{\colorbox{lightgreen}{\hspace{8pt}#1\hspace{8pt}}}
\newcommand{\tagno}[1]{\colorbox{lightred}{\hspace{8.5pt}#1\hspace{8.5pt}}}
\newcommand{\tagnop}[1]{\colorbox{lightred}{\hspace{6.8pt}#1\hspace{6.8pt}}}

\setlength{\fboxsep}{2pt}
\begin{tabular}{lccc}
    \toprule
     {\uppercaseaibot} & FP-Agent & Cloudflare & Rule Triggered \\
     \midrule
     Atlas Agent   & \tagyes{Yes} & \tagno{No}     & N/A \\
     Browser Use   & \tagyes{Yes} & \tagno{No}     & N/A \\
     Claude        & \tagyes{Yes} & \tagnop{No$^*$} & ClaudeBot \\
     Comet         & \tagyes{Yes} & \tagnop{No$^*$} & Perplexity-User \\
     Skyvern       & \tagyes{Yes} & \tagno{No}     & N/A \\
     ChatGPT Agent & \tagyes{Yes} & \tagnop{No$^*$} & ChatGPT-User \\
     Manus         & \tagyes{Yes} & \tagyes{Yes}   & Manus Bot \\
     \bottomrule
\end{tabular}

%% file: sections/07_conclusion.tex
\section{Conclusion}
\label{sec:conclusion}

In this paper, we presented \texttt{FP-Agent}, a measurement framework for characterizing and classifying browsing agents using browser and behavioral fingerprints.
Across seven browsing agents and human users, we showed that browser fingerprints alone provide limited discriminative power, especially when multiple agents share the same underlying browser fingerprint, whereas behavioral fingerprints offer stronger discriminative power.
In particular, browsing agents exhibit distinctive interaction-level behaviors---including differences in typing, scrolling, and mouse movement---that reliably separate them from humans and from one another.
Overall, \texttt{FP-Agent} produces novel insights that have strong implications for the AI bot detection arms race, providing an extensible approach for future research on {\aibot} identification.

%% file: sections/08_ethics.tex
\vspace{-0.5mm}
\section{Ethics}
\label{sec:ethics}

In this section, we discuss ethical considerations taken into account during this study.
Our university's Institutional Review Board (IRB) approved our human subjects experiment protocol by making a \textit{review exempt} determination.
Despite this, we uphold ethics around protecting participant privacy by creating a separate website version for human participants that hashes their IP addresses and disables scripts collecting identifying artifacts.
The participants were also compensated with extra credit for their time.
We strategically isolated the attribution mechanism for logging participants' task completion on a separate page of our website that does not contain our JavaScript instrumentation.
These completion records were deleted after awarding the extra credit.
We maintained transparency by sharing details about the data collection, handling, and usage before onboarding the participants. 
All included participants consented to be in the study and were allowed to stop participating at any point during the study.

We also ensured ethical compliance with {\aibot} providers during our study.
All {\aibot}s we tested either required a paid subscription or API credits from an LLM provider, meaning that all compute used for LLM reasoning was paid for.
We also respected the rate limits set by each {\aibot} provider when applicable.

%% file: sections/09_data_availability.tex
\section{Data Availability}
\label{appendix:data_availability}

To ensure reproducibility and foster future research efforts, we make \textit{all} artifacts of our work \textit{fully} available to the public.\\ 

\para{Code}
Code-related artifacts of our research can be accessed at \textit{\url{https://github.com/ethanbwang/fp-agent}}.
This includes our honey website infrastructure and data collection scripts comprised of \texttt{fingerprint.js} for browser fingerprint collection and a custom script for behavioral fingerprint collection as described in Section~\ref{sec:honey_website_design}, task implementations (discussed in Section~\ref{sec:tasks}), {\aibot} data collection automation (discussed in Section~\ref{sec:data_collection}),  as well as \texttt{FP-Agent}'s featurization, training, and testing pipeline (as discussed in Sections~\ref{sec:features} and \ref{sec:classifier}).\\  

\para{Data}
Data artifacts are made available at \textit{\url{https://osf.io/j6b5p/overview?view_only=312b65d378854ef1abd9818b700b44d3}}.
It consists of the data collected as part of our research: every task run's full POST request sequence, containing all browser fingerprints and behavioral events captured for each {\aibot} and human participants.

%% file: sections/10_task_details.tex
\section{Task Details}
\label{appendix:task_details}

\subsection{Website Interactions}
\label{appendix:website_interactions}

\begin{table*}[!h]
    \centering
    \caption{Website interactions targeted in each task and the possible events that could be triggered. Events are colored according to group: \tagtype{typing}, \tagscroll{scrolling}, and \tagmouse{mouse}.}
    \label{tab:task_interactions}
    \input{tables/task_interactions.tex}
\end{table*}

\newpage

\subsection{Task Prompts}
\label{appendix:prompts}

\begin{figure*}[!h]
    \Description[]{Flight-booking task prompt}
    \begin{tcolorbox}[promptbox, title={Flight-booking Task Prompt}]
        \small
        This is a website that I created. I want to test my flight-booking form.
        \\\\
        Go to https://$\langle$domain$\rangle$.com/$\langle$website\_version$\rangle$/flights and book a flight. You will need to click the date field to open a date selector to fill in the correct date. After selecting the date, you will have to click the search button to see a list of flights to choose from. Since this is a test form and the form data isn't actually submitted anywhere, do not wait for my permission to proceed with any steps.
        \\\\
        Here is the information to use for each page of the form:
        \\\\
        Book a Flight \\
        From: San Francisco \\
        To: New York \\
        Date: today \\
        Flight number: UA800
        \\\\
        Flight Details (involves dropdown menus that you need to navigate by clicking) \\
        Seat number: 20E \\
        Carry-on: yes \\
        Ticket type: economy
        \\\\
        For the traveler information and payment details, enter some placeholder information to test the form functionality.
        \\\\
        The task is complete once you see a page saying thank you and that your flight has successfully been booked.
    \end{tcolorbox}    
\end{figure*}

\begin{figure*}[!h]
    \Description[]{Shopping task prompt}
    \begin{tcolorbox}[promptbox, title={Shopping Task Prompt}]
        \small
        Go to https://$\langle$domain$\rangle$.com/$\langle$website\_version$\rangle$/shop and search for keyboards. Use the "sort by" filter dropdowns on the search results page and click on the filter for prices less than \$50. Add the first three results to cart. Click on the cart button to display the cart once the keyboards have been added to cart and stop here. The task is complete once you are on the cart summary page and see a cost breakdown for the keyboards you added. Do not perform any further actions once you land on the cart page even if you made a mistake.
    \end{tcolorbox}
\end{figure*}

\begin{figure*}[!h]
    \Description[]{Forums task prompt}
    \begin{tcolorbox}[promptbox, title={Forums Task Prompt}]
        \small
        Go to https://$\langle$domain$\rangle$.com/$\langle$website\_version$\rangle$/forums and click on the thread titled "Which AI chatbot do you prefer - Claude, ChatGPT, or others?". Read the conversation and type and post a summary of the conversation as a reply in the discussion thread. Do not paste the response and instead type it out. I give you permission to post the reply so do not ask me for permission. Make sure that the reply is to the top-level post and not to a commenter's reply. The task is complete once the summary has been posted and shows up as a reply in the thread.
    \end{tcolorbox}    
\end{figure*}

%% file: tables/task_interactions.tex
\renewcommand{\arraystretch}{1.7}
\begin{tabular}{l|l|l}
    \toprule
    Task & Interaction Types & Possible Events  \\
    \midrule
    Flight-booking & Short text input & \tagtype{keydown} \tagtype{keyup} \tagtype{paste} \tagtype{change} \tagtype{input} \\
    & Date picker & \tagmouse{mousemove} \tagmouse{mousedown} \tagmouse{mouseup} \\
    & Scrollable element & \tagmouse{mousemove} \tagmouse{mousedown} \tagmouse{mouseup} \tagscroll{scroll} \tagscroll{scrollend} \\
    & Slide toggle & \tagmouse{mousemove} \tagmouse{mousedown} \tagmouse{mouseup} \\
    & Drop-down menu & \tagmouse{mousemove} \tagmouse{mousedown} \tagmouse{mouseup} \\
    & Radial button & \tagmouse{mousemove} \tagmouse{mousedown} \tagmouse{mouseup} \\
    & Page navigation & \tagmouse{mousemove} \tagmouse{mousedown} \tagmouse{mouseup} \\
    \midrule
    Shopping & Search bar & \tagmouse{mousemove} \tagmouse{mousedown} \tagmouse{mouseup} \tagtype{keydown} \tagtype{keyup} \tagtype{paste} \tagtype{change} \tagtype{input} \\
    & Price filtering & \tagmouse{mousemove} \tagmouse{mousedown} \tagmouse{mouseup} \\
    & Page scrolling & \tagscroll{scroll} \tagscroll{scrollend} \\
    & Page navigation & \tagmouse{mousemove} \tagmouse{mousedown} \tagmouse{mouseup} \\
    \midrule
    Forums & Page navigation & \tagmouse{mousemove} \tagmouse{mousedown} \tagmouse{mouseup} \\
    & Page scrolling & \tagscroll{scroll} \tagscroll{scrollend} \\
    & Long text input & \tagtype{keydown} \tagtype{keyup} \tagtype{paste} \tagtype{change} \tagtype{input} \\
    \bottomrule
\end{tabular}
\renewcommand{\arraystretch}{1.0}

%% file: sections/11_task_screenshots.tex
\section{Screenshots of Tasks}
\label{appendix:task_screenshots}

\begin{figure*}[!ht]
    \centering
    \includegraphics[width=1\linewidth]{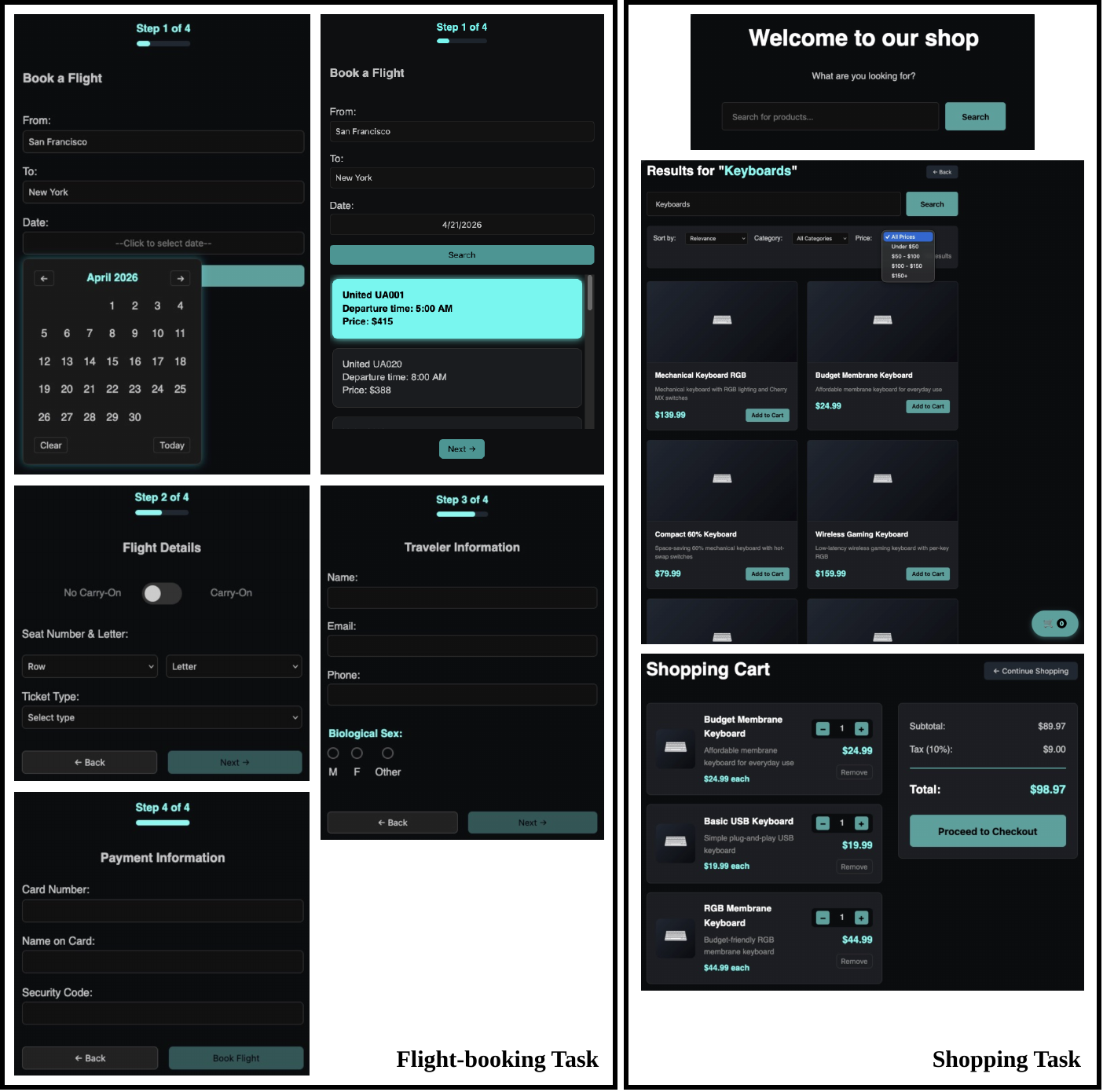}
    \caption{Flight-booking task (left) and shopping task (right).}
    \Description[]{Flight-booking task (left) and shopping task (right).}
    \label{fig:flight_booking_and_shopping_screenshots}
\end{figure*}

\begin{figure*}[!ht]
    \centering
    \includegraphics[width=1\linewidth]{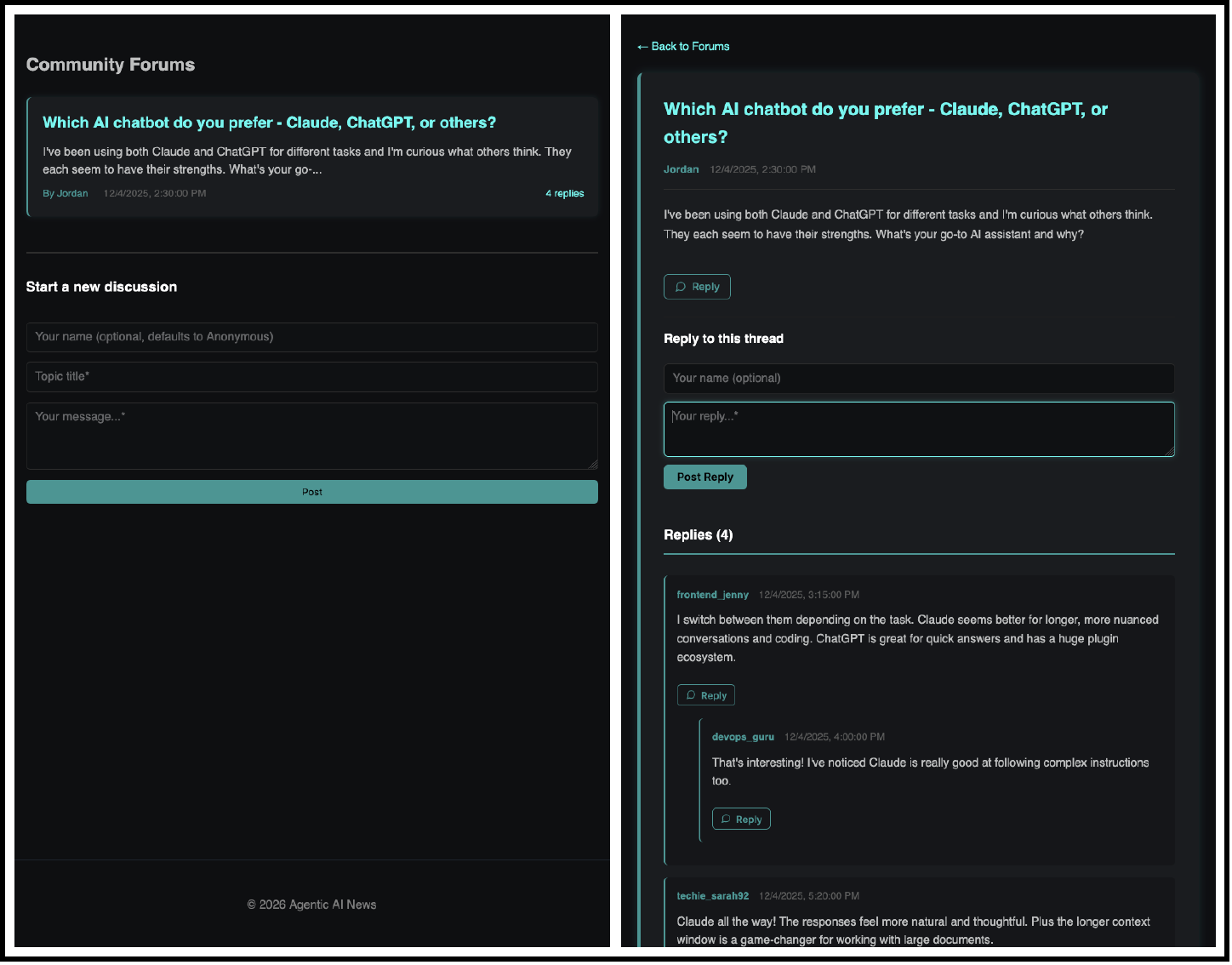}
    \caption{Forums task.}
    \Description[]{Forums task.}
    \label{fig:forums_screenshots}
\end{figure*}

%% file: sections/12_behavioral_features.tex
\section{Behavioral Features}
\label{appendix:behavioral_features}

\begin{table*}[!ht]
    \centering
    \fontsize{9}{10.5}\selectfont
    \caption{List of behavioral features. Mouse button down/up ratio measures the ratio of \texttt{mousedown} to \texttt{mouseup} events. Ranks are from the classifier using the combined feature set classifying all classes. Dashes indicate zero importance.}
    \label{tab:behavioral_feature_table}
    \input{tables/behavioral_feature_table}
\end{table*}

%% file: tables/behavioral_feature_table.tex
\begin{tabular}{c|lcc}
    \toprule
    & Feature & Type & Importance Rank \\
    \midrule
    \multirow{24}{*}{\rotatebox{90}{Mouse}} & Mouse movement angle of curvature range & Numeric & 2 \\
    & Number of mouse & Numeric & 11 \\
    & Presence of mouse button 0 & Boolean & 13 \\
    & Mouse movement angle of curvature mean & Numeric & 18 \\
    & Mouse movement direction mean & Numeric & 22 \\
    & Mouse button 0 down/up ratio & Numeric & 24 \\
    & Mouse movement angle of curvature standard deviation & Numeric & 25 \\
    & Mouse movement curvature distance median & Numeric & 26 \\
    & Mouse movement direction range & Numeric & 27 \\
    & Mouse movement curvature distance mean & Numeric & 33 \\
    & Presence of mouse move events & Boolean & 34 \\
    & Mouse movement direction standard deviation & Numeric & 36 \\
    & Mouse movement direction median & Numeric & - \\
    & Mouse movement angle of curvature median & Numeric & - \\
    & Mouse movement curvature distance range & Numeric & - \\
    & Mouse movement curvature distance standard deviation & Numeric & - \\
    & Presence of mouse button 1 & Boolean & - \\
    & Mouse button 1 down/up ratio & Numeric & - \\
    & Presence of mouse button 2 & Boolean & - \\
    & Mouse button 2 down/up ratio & Numeric & - \\
    & Presence of mouse button 3 & Boolean & - \\
    & Mouse button 3 down/up ratio & Numeric & - \\
    & Presence of mouse button 4 & Boolean & - \\
    & Mouse button 4 down/up ratio & Numeric & - \\
    \midrule
    \multirow{16}{*}{\rotatebox{90}{Typing}} & Presence of paste event & Boolean & 1 \\
    & Hold latency median & Numeric & 3 \\
    & Inter-key latency median & Numeric & 4 \\
    & Hold latency mean & Numeric & 5 \\
    & Number of change events & Numeric & 7 \\
    & Number of input events & Numeric & 10 \\
    & Hold latency range & Numeric & 14 \\
    & Inter-key latency mean & Numeric & 15 \\
    & Dangling keydown event (keydown not paired with keyup) & Boolean & 19 \\
    & Inter-key latency range & Numeric & 23 \\
    & Hold latency standard deviation & Numeric & 29 \\
    & Inter-key latency standard deviation & Numeric & 31 \\
    & Number of backspace/delete keypresses & Numeric & 32 \\
    & Presence of keypresses (keydown and keyup) & Boolean & 35 \\
    & Dangling keyup event (keyup not paired with keydown) & Boolean & - \\
    & Ratio of backspace/delete to total keypresses & Numeric & - \\
    \midrule
    \multirow{10}{*}{\rotatebox{90}{Scrolling}} & Scroll distance standard deviation & Numeric & 6 \\
    & Scroll distance mean & Numeric & 8 \\
    & Scroll time median & Numeric & 9 \\
    & Scroll distance range & Numeric & 12 \\
    & Scroll time mean & Numeric & 16 \\
    & Scroll distance median & Numeric & 17 \\
    & Scroll time standard deviation & Numeric & 20 \\
    & Presence of scroll event & Boolean & 21 \\
    & Scroll time range & Numeric & 28 \\
    & Presence of scroll end event & Boolean & 30 \\
    \bottomrule
\end{tabular}

%% file: sections/13_confusion_matrix.tex
\section{FP-Agent Confusion Matrices}
\label{appendix:fpagent_confusion_matrices}

\begin{figure*}[!h]
    \centering
    \includegraphics[width=1\linewidth]{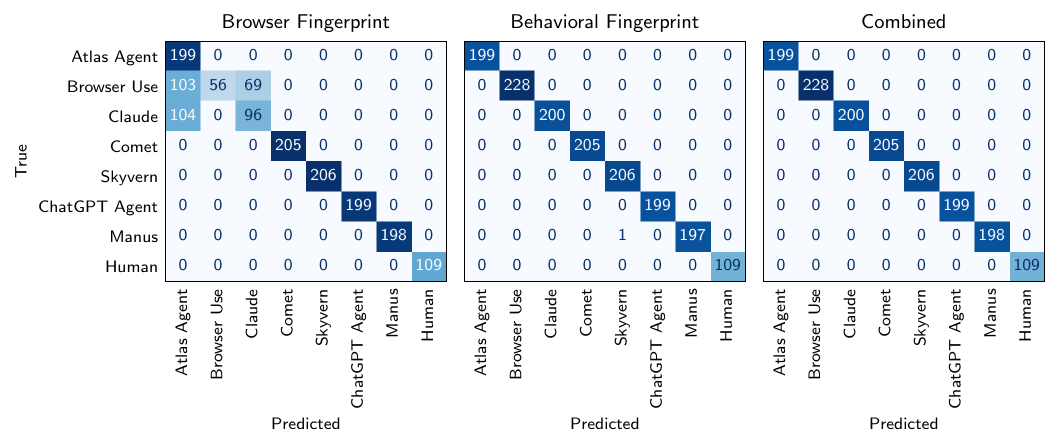}
    \caption{\texttt{FP-Agent's} confusion matrices on each feature set across all classes. The behavioral- and combined-fingerprint classifiers perform almost perfectly, whereas the duality of perfect classification for classes not sharing fingerprints and imperfect classification for classes sharing fingerprints suggests the browser-fingerprint classifier is overfitting, providing additional evidence that browser fingerprints alone are insufficient for our classification problem.
    }
    \Description[]{Confusion matrices for all classifiers.}
    \label{fig:confusion_matrices}
\end{figure*}

%% file: sections/14_keystroke_dynamics.tex
\section{Keystroke Latency Statistics}
\label{appendix:keystroke_latency_stats}

\begin{table*}[!h]
    \centering
    \small
    \caption{Keystroke dynamics statistics by class. The 95\% confidence intervals (CI) for the means are in the square brackets.}
    \label{tab:keystroke_stats}
    \input{tables/average_interkey_and_hold_latencies}
\end{table*}

%% file: tables/average_interkey_and_hold_latencies.tex
\begin{tabular}{lcccccc}
    \toprule
     & \multicolumn{3}{c}{Interkey Latency (ms)} & \multicolumn{3}{c}{Hold Latency (ms)} \\
    \cmidrule(lr){2-4} \cmidrule(lr){5-7}
    Class & Mean & 95\% CI & Standard Deviation & Mean & 95\% CI & Standard Deviation \\
    \midrule
    Atlas Agent & N/A & N/A & N/A & N/A & N/A & N/A \\
    Browser Use & 5.31 & [4.99, 5.63] & 0.19 & 10.19 & [9.92, 10.46] & 0.90 \\
    Claude & 0.56 & [0.52, 0.59] & 0.17 & 0.94 & [0.88, 1.00] & 0.24 \\
    Comet & N/A & N/A & N/A & 2.97 & [2.85, 3.08] & 1.66 \\
    Skyvern & 9.52 & [9.48, 9.56] & 0.81 & 11.33 & [11.30, 11.35] & 0.56 \\
    ChatGPT Agent & N/A & N/A & N/A & 66.58 & [66.55, 66.62] & 34.46 \\
    Manus & 1.39 & [1.37, 1.41] & 0.21 & 52.92 & [52.88, 52.95] & 0.53 \\
    Human & 120.43 & [114.97, 125.89] & 78.74 & 97.48 & [95.47, 99.50] & 27.13 \\
    \bottomrule
\end{tabular}